\documentclass[twocolumn]{aastex631}

\usepackage{amsmath}	

\newcommand{\Ctsz}{C^{yy}_{\ell}}
\newcommand{\Cgal}{C^{\rm gg}_{\ell}}
\newcommand{\Ccross}{C^{\text{g}y}_{\ell}}
\newcommand{\elleff}{\ell_{\rm eff}}

\usepackage[utf8x]{inputenc}

\shorttitle{tSZ and projected density cross-correlation}
\shortauthors{Ibitoye et al.}

\graphicspath{{./}{figures/}}

\begin{document}

\title{Cross Correlation between thermal Sunyaev-Zel'dovich effect and projected galaxy density field }

\author[0000-0002-0966-8598]{Ayodeji Ibitoye}
\affiliation{School of Chemistry and Physics, University of KwaZulu-Natal, Westville Campus \\
Private Bag X54001, Durban, 4000, South Africa}
\affiliation{ NAOC-UKZN Computational Astrophysics Centre (NUCAC) \\
University of KwaZulu-Natal, Durban, 4000, South Africa}
\affiliation{Department of Physics and Electronics, Adekunle Ajasin University \\
P. M. B. 001, Akungba-Akoko, Ondo State, Nigeria}

\author{Denis Tramonte}
\affiliation{Purple Mountain Observatory, CAS, No.10 Yuanhua Road\\
Qixia District, Nanjing 210034, China}
\affiliation{School of Chemistry and Physics, University of KwaZulu-Natal, Westville Campus \\
Private Bag X54001, Durban, 4000, South Africa}
\affiliation{ NAOC-UKZN Computational Astrophysics Centre (NUCAC) \\
University of KwaZulu-Natal, Durban, 4000, South Africa}

\author[0000-0001-8108-0986]{Yin-Zhe Ma$^{\dagger}$}
\affiliation{School of Chemistry and Physics, University of KwaZulu-Natal, Westville Campus \\
Private Bag X54001, Durban, 4000, South Africa}
\email{{\,}$^{\dagger}$Corresponding author: Y.-Z. Ma, ma@ukzn.ac.za}
\affiliation{ NAOC-UKZN Computational Astrophysics Centre (NUCAC) \\
University of KwaZulu-Natal, Durban, 4000, South Africa}
\affiliation{Purple Mountain Observatory, CAS, No.10 Yuanhua Road\\
Qixia District, Nanjing 210034, China}

\author{Wei-Ming Dai}
\affiliation{School of Chemistry and Physics, University of KwaZulu-Natal, Westville Campus \\
Private Bag X54001, Durban, 4000, South Africa}
\affiliation{ NAOC-UKZN Computational Astrophysics Centre (NUCAC) \\
University of KwaZulu-Natal, Durban, 4000, South Africa}

\begin{abstract}

We present a joint analysis of the power spectra of the \textit{Planck} Compton $y$-parameter map and the projected galaxy density 
field using the Wide Field Infrared Survey Explorer (WISE) all-sky survey. We detect the statistical correlation between WISE and \textit{Planck} data (g$y$) with a significance 
of $21.8\,\sigma$. We also measure the auto-correlation spectrum for the tSZ ($yy$) and the galaxy density field maps (gg) with a significance 
of $150\,\sigma$ and $88\,\sigma$, respectively. We then construct a halo model and use the measured correlations $C^{\rm gg}_{\ell}$, $C^{yy}_{\ell}$ 
and $C^{{\rm g}y}_{\ell}$ to constrain the tSZ mass bias $B\equiv M_{500}/M^{\rm tSZ}_{500}$. We also fit for the galaxy bias, which is 
included with explicit redshift and multipole dependencies as $b_{\rm g}(z,\ell)=b_{\rm g}^0(1+z)^{\alpha}(\ell/\ell_0)^{\beta}$, with $\ell_0=117$. 
We obtain the constraints to be $B =1.50{\pm 0.07}\,(\textrm{stat}) \pm{0.34}\,(\textrm{sys})$, i.e. 
$1-b_{\rm H}=0.67\pm 0.03\,({\rm stat})\pm 0.16\,({\rm sys})$ (68\% confidence level) for the hydrostatic mass bias, and  $b_{\rm g}^0=1.28^{+0.03}_{-0.04}\,(\textrm{stat}) \pm{0.11}\,(\textrm{sys})$, with $\alpha=0.20^{+0.11}_{-0.07}\,(\textrm{stat}) \pm{0.10}\,(\textrm{sys})$ and $\beta=0.45{\pm 0.01}\,(\textrm{stat}) \pm{0.02}\,(\textrm{sys})$ for the galaxy bias. 
Incoming data sets from future CMB and galaxy surveys (e.g. Rubin Observatory) will allow probing the large-scale gas 
distribution in more detail.
\end{abstract}

\keywords{galaxies: clusters: general - galaxies: clusters: intracluster medium - intergalactic medium - cosmology: large-scale structure of Universe}


\section{Introduction} 
\label{sec:intro}
The thermal Sunyaev-Zeldovich (tSZ) effect is a secondary anisotropy of the cosmic microwave background 
(CMB) radiation caused by the inverse Compton scattering of CMB photons off warm-hot electrons. This 
phenomenon results in an effective CMB spectral distortion which can be quantified as~\citep{carlstrom02}:
\begin{equation}
\label{eq:tsz}
\frac{\Delta T}{T_{\rm CMB}}= g(x)y,
\end{equation}
where $T_{\rm CMB}= 2.725\,{\rm K}$ is the mean CMB temperature, $y$ is the Compton parameter and the 
function $g(x)$ quantifies the frequency dependence of the tSZ effect. The latter is expressed as a 
function of the dimensionless frequency $x\equiv h \nu/{ k_{\rm B}T_{\rm CMB}}$, with $h$ the Planck 
constant, $\nu$ the photon frequency and $k_{\rm B}$ the Boltzmann constant. Neglecting relativistic 
corrections, the function $g(x)$ reads:
\begin{equation}
	g(x) = x\coth{\left(\frac{x}{2} \right)} - 4.
\end{equation}
The Compton parameter $y$ quantifies the amplitude of the tSZ effect independently of the observing 
frequency. It is proportional to the electron pressure integrated along the line-of-sight (LoS) distance $l$:
\begin{equation}
\label{eq:compton}	
y=\frac{\sigma_{\rm T}k_{\rm B}}{m_{\rm e}c^{2}}\int n_{\rm e}T_{\rm e}\,{\rm d}l,
\end{equation}
where $n_{\rm e}$ and $T_{\rm e}$ are the electron number density and temperature, $\sigma_{\rm T}$ 
is the Compton cross section and $m_{\rm e}$ is the electron mass. Low-mass and high-$z$ clusters, 
for which an individual detection is generally difficult, provide a significant integrated contribution 
to $y$ which is detectable by measuring the angular power spectrum of the tSZ effect, particularly 
at small scales~\citep{Trac2011,P11}. The tSZ angular power spectrum is then an excellent probe  
of the physical conditions of the hot gas around dark matter haloes~\citep{Seljak2002}.

A key feature of the tSZ is that it is not explicitly dependent on redshift; the LoS integral in 
Eq.~(\ref{eq:compton}) implies that all of the warm-hot gas encountered by CMB photons from the 
last-scattering surface up to the observer contributes to the spectral distortion. In this context, 
cross-correlating the observed tSZ with other large-scale structure (LSS) tracers is a very useful tool 
to recover information on the redshift of the responsible hot gas; this, in turn, allows for a   
better characterisation of the diffuse gas component distribution in relation with the cosmic web, 
eventually providing insights into the growth of structures. Such LSS tracers are usually provided 
by optical survey measurements, and many works in recent years have contributed to their 
exploitation in this sense. This type of cross-correlation analysis has been conducted using 
galaxy clustering~\citep{Pandey19,Nick2020,chiang20}, 
weak lensing~\citep{V.Waerbeke2014,Hill2014,Ma2015,Atrio-Barandela2017}, cosmic shear~\citep{Hojjati2017}, 
luminous red galaxies~\citep{tanimura19,de-Graaff2019}, cosmic voids~\citep{Alonso2018}, galaxy 
groups~\citep{Hill2018, lim20} and galaxy clusters~\citep{Kita_and_komatsu,Bolliet18,Bolliet2020,rotti21}. 

One of the key parameters entering SZ-based studies of the gravitational clustering of dark 
matter haloes is the tSZ mass bias, which is defined as the ratio~\citep{Planck2016}:
\begin{equation}
	\label{eq:biasdef}
B = \frac{M_{500}}{M^{\rm tSZ}_{500}},
\end{equation} 
whereas in some literature it is inversely defined as $1-b_{\rm H}\equiv M^{\rm tSZ}_{500}/M_{500}$. 
In Eq.~(\ref{eq:biasdef}) $M_{500}$ is the cluster overdensity mass defined with respect to 
the Universe critical density at that redshift (see also Sec.~\ref{ssec:halomodel} for details), 
and represents the ``true'' cluster mass. The quantity $M^{\rm tSZ}_{500}$ is instead the cluster 
mass inferred from the measured tSZ flux. Any systematics affecting the mass measurement is then 
encoded in the bias parameter $B$. The main contribution to $B$ is most likely the assumption of 
hydrostatic equilibrium in the intra-cluster medium, which is made in the modeling of the ICM 
pressure profile~\citep{Arnaud2010}, but not necessarily satisfied by the detected clusters. To 
this, we can add the contribution of instrumental calibration and additional systematics in the 
underlying X-ray modeling, which is required to provide mass proxies and calibrate the mass 
estimation~\citep{Planck2015-XXIV}. The current uncertainty on the mass bias is one of the major 
issues hindering the full exploitation of cluster-related observables as cosmological tools. 

Several studies have recently conducted cross-correlation studies between the tSZ and other tracers 
to constrain the tSZ mass bias. For example, \citet{Nick2020} cross-correlated the tSZ maps from 
{\it Planck\,} with the projected galaxies sourced from a combination of the near-infrared 2 Micron 
All-Sky Survey (2MASS), optical SuperCOSMOS, and the mid-infrared Wide Field Infrared Survey 
Explorer~\citep[WISE,][]{WISE2010} at $z \lesssim 0.4$, and achieved the constraint 
$1-b_{\rm H}=0.59\pm0.03$.~\citet{Makiya2019} used the tSZ map from {\it Planck} and the 2MASS redshift 
survey (2MRS) catalogue to the same aim, constraining the bias to be $1-b_{\rm H}=0.649\pm0.041$. Similar 
studies can be found in~\citet{Hurier2017},~\citet{Bolliet18},~\citet{Salvati-bias},~\citet{Ken-Osato2019} 
and~\citet{Zubeldia2019}. These findings are summarised in Table~\ref{tab:Bias_comparison}. Our study 
pursues a similar scientific goal, but for the first time utilising uniquely the all-sky WISE galaxy 
catalogue in combination with \textit{Planck} tSZ maps. More precisely, we calculate the cross-correlation 
spectrum between the {\it Planck} tSZ map and the projected galaxy density field map obtained from the 
WISE catalogue, and the auto-correlation spectrum for each observable. We then employ a halo model 
framework to theoretically predict all three correlation cases, and fit for the tSZ mass bias by 
jointly comparing the predicted spectra with our measurements. Our analysis will also allow us to 
place novel constraints on the galaxy bias, which is a fundamental ingredient to model the projected 
galaxy field, as well as on other parameters quantifying the foreground contaminations affecting the tSZ map. 
 
This paper is organized as follows. In Section~\ref{sec:data} we describe the data set we use. The 
measurements of the auto- and cross-correlations are presented in Section~\ref{sec:analysis}, while 
Section~\ref{sec:modelling} details their theoretical predictions in a halo model framework. In 
Section~\ref{sec:parest}, we present the methodology and results of our parameter estimation, 
and discuss the resultant implications. The concluding remarks are presented in Section~\ref{sec:conclusions}. 
Throughout this work we assume a spatially flat $\Lambda$-CDM cosmology with cosmological parameters 
fixed to {\it Planck\,} 2018 best-fitting values, i.e. $\Omega_{\rm c}h^{2}=0.120$, $\Omega_{\rm b}h^{2}=0.0223$, 
$h=0.674$, $n_{\rm s}=0.965$, $\tau=0.0540$ and $\ln(10^{10}A_{\rm s})=3.043$~\citep{planck18}. 

\section{Data}
\label{sec:data}

In the analysis of this paper we employ the Compton-$y$ map from the \textit{Planck} 2015 data 
release~\citep{Planck2016_sz} and the projected galaxy density field from the WISE All Sky 
Catalogue~\citep{WISE2010}. A detailed description of this data set is provided in the following.


\subsection{Compton parameter map} 
\label{ssec:planck}
We use the full sky Compton parameter map issued by the \textit{Planck} Collaboration~\citep{Planck2016_sz}. 
The map was generated by a combination of \textit{Planck} individual frequency maps with tailored 
algorithms that enhance the SZ signal and suppress the contribution from the CMB and other Galactic 
and extra-Galactic foregrounds. The individual frequency maps were convolved to a common resolution of 
$10'$, which also sets the resolution of the resulting $y$ map. We stress that the latter is still 
affected by foreground residuals, particularly by thermal dust emission at large scales and clustered 
cosmic infrared background (CIB) and Poisson-distributed radio and infrared sources at small 
scales~\citep{Planck2014_tSZ,Planck2016_sz}. These spurious contaminants will be accounted for in our 
modeling analysis in Sec.~\ref{ssec:foregrounds}.

These data are publicly available and can be downloaded from the \textit{Planck} Legacy 
Archive\footnote{\url{http://pla.esac.esa.int/pla}}. The legacy products provide two independent all-sky 
maps produced using different linear combination algorithms, namely the Needlet Independent Linear Combination 
(NILC) method~\citep{remazeilles11} and the Modified Internal Linear Combination Algorithm~\citep[MILCA, ][]{hurier13}. 
The results presented in~\citet{Planck2016_sz} suggest that, at multipoles $\ell < 30$, the amplitude of the 
tSZ power spectrum measured on the MILCA-reconstructed $y$ map is slightly higher than the one measured on the 
NILC-reconstructed $y$ map. This difference is ascribed to a higher degree of contamination from thermal dust 
emission at large scales in the MILCA map. Hence, we kept the NILC map as our preferential choice and will present 
the results for this $y$ map only. We remind, however, that for the remaining part of the explored multipole range, 
the power spectra extracted from the two versions of the $y$ map prove consistency with~\citet{Planck2016_sz}, and we 
did check that, in our power spectrum measurement, results obtained using the two maps agree within their uncertainties. 

The maps are delivered in {\tt HEALPix} format~\citep{Gorski2005} with an original pixel resolution set 
by the parameter $N_{\rm side}=2048$. To optimise the efficiency of our data processing pipeline, the 
pixelisation was degraded to a lower $N_{\rm side}=512$, which is sufficient for our purpose. To suppress 
the contribution from Galactic foregrounds, we finally impose a 40\% Galactic plane mask, which is also 
available in the \textit{Planck} Legacy Archive. 

\begin{figure}
	\centering
	\includegraphics[width=8cm]{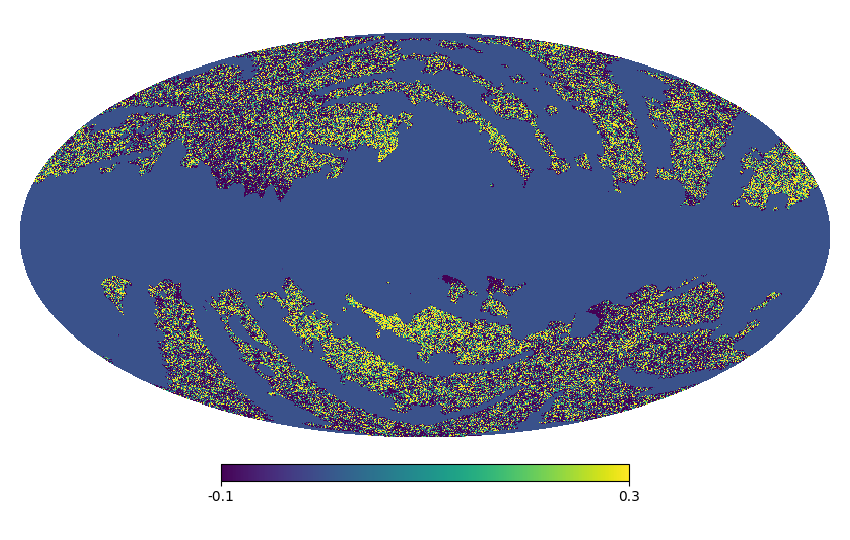}
	\caption{WISE galaxy overdensity map, computed using Eq.~(\ref{eq:wisemap}). 
	The masked region, where the overdensity is null, includes the contribution of 
	both the Galactic plane mask and the cuts applied to the WISE catalogue to remove 
	pointings affected by Moon contamination.}
	\label{fig:wisemap}
\end{figure}

\subsection{Galaxy overdensity maps}
\label{ssec:wise}

The WISE satellite scanned the whole sky in four photometric bands at 3.4, 4.6, 12 and $22\;\mu\text{m}$ 
(labelled as bands W1 to W4). The survey had enhanced sensitivity and angular resolution compared to 
previous infrared missions such as the InfraRed Astronomical Satellite~\citep[\textit{IRAS}][]{neugebauer84} 
or the Two Micron All-Sky Survey~\citep[2MASS,][]{skrutskie06}. In the context of this paper WISE represents 
therefore the ideal candidate to map the distribution of galaxies over the full sky. 

The resulting WISE All-Sky Data were made publicly available in 2012 and can be accessed at the NASA/IPAC 
Infrared Science Archive\footnote{\url{https://irsa.ipac.caltech.edu/cgi-bin/Gator/nph-scan?mission=irsa&submit=Select&projshort=WISE}}. 
In this paper, we employed the WISE All-Sky Source Catalogue~\citep{WISE2010,cutri12} which contains 
positional and photometric data for more than 563 million sources detected at more than $5\sigma$ in at least 
one band. Among these are Galactic stars, galaxies and quasars, plus other unidentified sources. Hence, the 
catalogue needs to be suitably queried to extract the galactic objects representing the targets of our analysis. 

The selection is performed based on the source flux values across different wavelengths, as bands W1 and 
W2 are mainly sensitive to Galactic or extra-Galactic starlight, while bands W3 and W4 probe thermal dust 
emission from the interstellar medium. Specifically, we follow the criteria outlined in~\citet{Ferraro2015}: 
we first apply the cut W1 $< 16.6$, which ensures a 95\% completeness of the resulting catalogue. According 
to the same reference, this cut also ensures substantial sample uniformity on the sky at high galactic latitudes, 
despite the WISE inhomogeneous scanning strategy. Second, we consider sources satisfying the condition 
W1 - W2 $> 0$, which is typically found in galaxies. We then make use of additional flags to remove 
contaminations and spurious signals. Pointings close to the Moon are affected by its infrared emission, 
the effect being quantified by the field \texttt{moon\_lev}; we mitigate the Moon contamination by discarding 
all sources for which \texttt{moon\_lev}$>4$. Finally, artifacts are eliminated by selecting only sources 
with the associated field \texttt{cc\_flag}$=0$. The resulting queried catalogue consists of 50,030,431 sources, 
which are the galaxies we employ to reconstruct the matter overdensity field. 

The galaxy number density map is generated by projecting the object catalogue onto an {\tt HEALPix} map with 
resolution $N_{\rm side}=512$. To the map we overlay the same Galactic plane mask used for the Compton parameter 
map, which, combined with the queries we applied on the WISE catalogue, yields a final unmasked sky fraction of 
$f^{\rm gg}_{\rm sky}=0.40$. The resulting mean number of galaxies per pixel (computed outside the masked area) 
is $\bar{n}_{\rm g} \simeq 39.96$. If we denote by $n_{\rm g}$ the number of galaxies in a generic pixel, then 
the corresponding galaxy density fluctuation $\delta_{\rm g}$ for that pixel is computed as:
\begin{equation}
\label{eq:wisemap}
\delta_{\rm g}=  \frac{n_{\rm g} - \bar{n}_{\rm g}}{\bar{n}_{\rm g}}.
\end{equation}  
Fig.~\ref{fig:wisemap} shows the resultant WISE galaxy overdensity map, where the masked region covers the 
Galactic plane plus a series of stripes that result from the excision of Moon contaminated pointings. 

\begin{figure}
	\centering
	\includegraphics[width= 8cm]{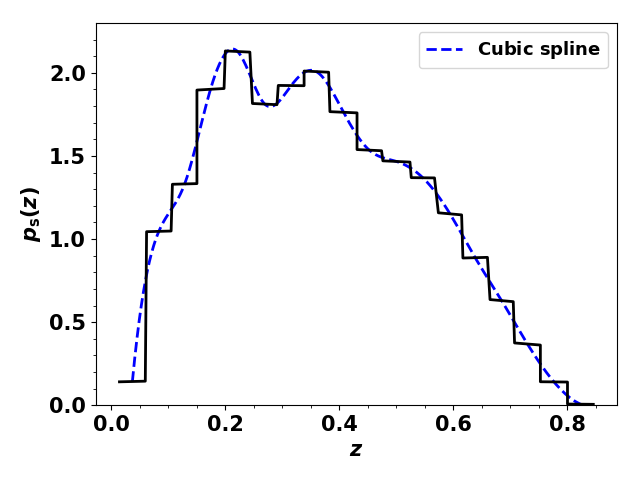}
	\caption{Histogram showing the normalised redshift distribution $p_{\rm s}(z)$ of 
	WISE galaxies, taken from Fig. 4 in~\citet{Yan2013}. The fitting function is plotted as the dashed blue line.}
\label{fig:dndz-fit}
\end{figure}

WISE photometric data cannot yield direct estimates of the object redshifts. The analysis conducted in this 
paper, however, does not require the knowledge of individual source redshifts, as only the redshift 
distribution of the galaxy number density $p_{\rm s}(z)$ will be needed in our theoretical modeling 
(Sec.~\ref{ssec:gfourier}). To this aim, we adopted the statistical distribution of galaxies derived 
in~\cite{Yan2013} via cross-matching with SDSS DR7 data~\citep{abazajian09}. The distribution is plotted in 
Fig.~\ref{fig:dndz-fit} and is found to peak at $z\sim0.24$, spanning the range from $z=0$ to $z\sim0.85$. 
For the subsequent theoretical modeling described in Sec.~\ref{sec:modelling} it is useful to have the redshift 
distribution parametrised analytically, which allows to evaluate the expected galaxy number density for any 
value of redshift within the covered range. For this purpose, we employ the \texttt{Python Scipy} cubic 
spline, which we found can reproduce the features of the histogram in Fig.~\ref{fig:dndz-fit} with higher fidelity compared 
to a polynomial fitting.


\section{Measurements of angular power spectra}
\label{sec:analysis}

\begin{figure}
\centering
\includegraphics[width=8cm]{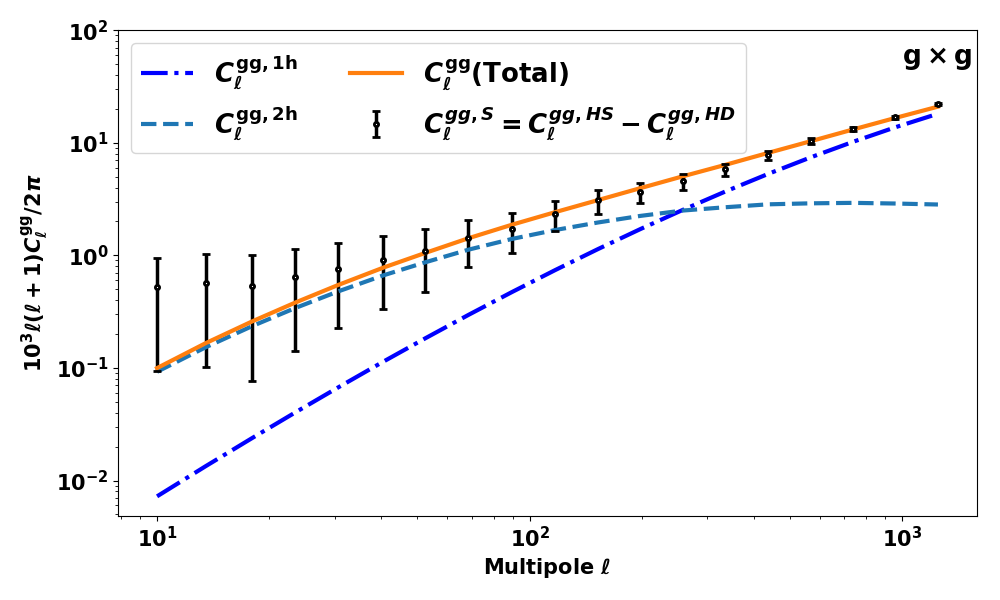}
\includegraphics[width=8cm]{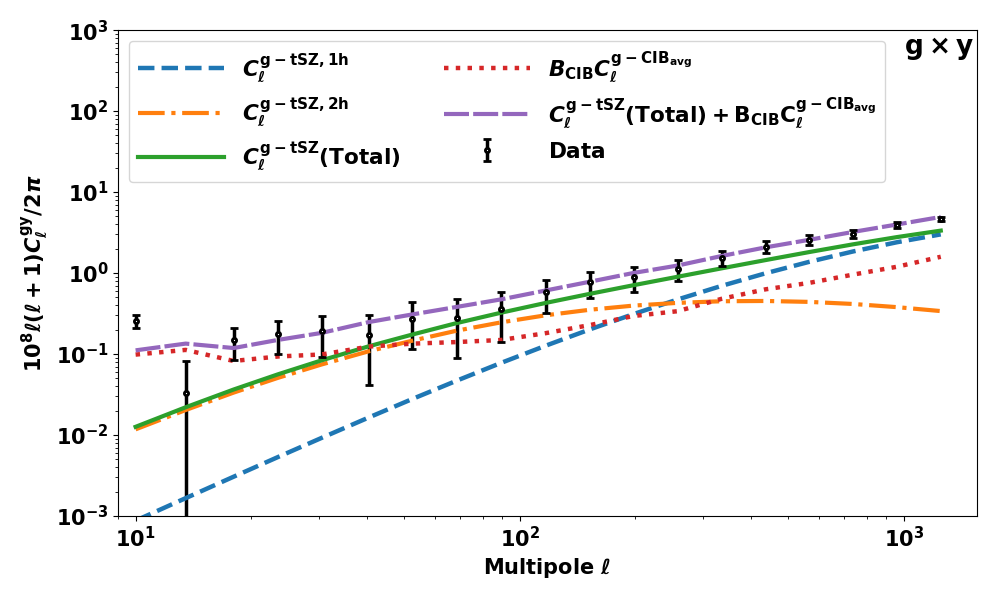}
\includegraphics[width=8cm]{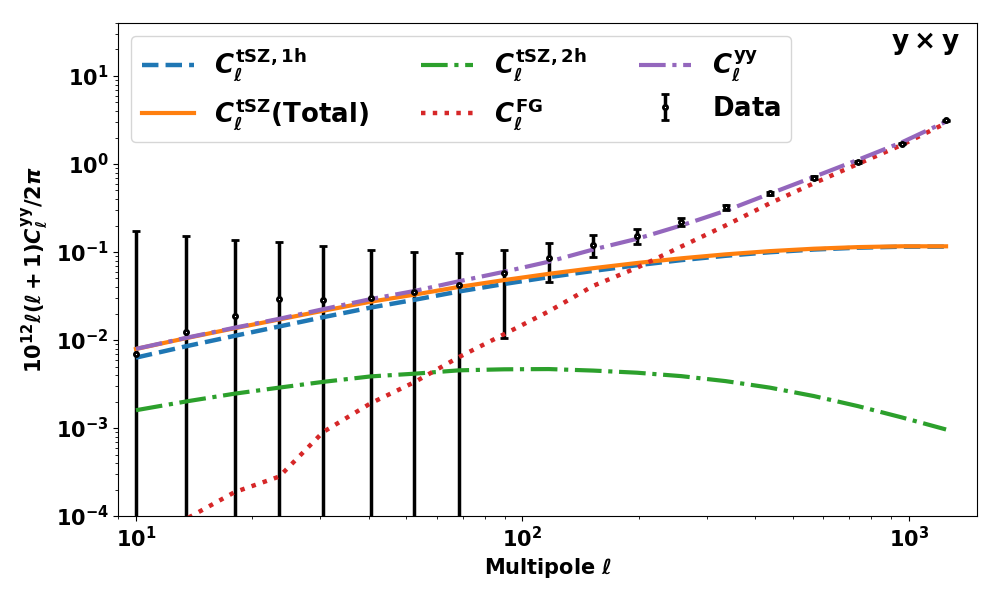}
\caption{Power spectra for the gg (top), g$y$ (middle) 
	 and $yy$ (bottom) correlations. Points represent our measurements binned 
	 at the effective multipoles from Table~\ref{tab:spectra}, with error bars computed 
	 as the square root of the diagonal terms in the corresponding covariance 
	 matrices. Lines represent the associated theoretical predictions computed 
	 using the best-fit parameters quoted in Table~\ref{tab:estimates} (full covariance case),
	 obtained as described in Sec.~\ref{sec:parest}. For each case we 
	 show separately the contribution 
	 of the one-halo and the two-halo terms, and the contribution of the foregrounds
	 affecting the correlation. }
\label{fig:spectra}	
\end{figure}

The Compton parameter map and the projected galaxy overdensity map described in Secs.~\ref{ssec:planck} 
and~\ref{ssec:wise} are used to measure the auto-correlation angular power spectrum of each observable, 
and the cross-correlation angular power spectrum between the two. To this aim, we employ the Spatially 
Inhomogeneous Correlation Estimator for Temperature and 
Polarisation\footnote{\url{http://www2.iap.fr/users/hivon/software/PolSpice/}.}~\citep[\texttt{PolSpice}, ][]{challinor11} 
package, which is a tool to statistically analyze any data pixelated over a sphere. The software accepts 
as input any combination of maps in {\tt HEALPix} format, together with a possible sky mask or pixel weighting 
scheme, and it delivers as output the corresponding auto- or cross-power spectrum or two-point correlation 
function. For our purpose, it is more convenient to work with power spectra, so we will not be using correlation 
functions in this paper. 

By inputting the maps described in Sec.~\ref{sec:data} and the associated masks to \texttt{PolSpice}, 
we obtain the resulting tSZ angular power spectrum $\Ctsz$, the power spectrum of the galaxy overdensity 
field $\Cgal$ and the cross-correlation power spectrum $\Ccross$. Hereafter we shall label them as the 
$yy$, gg and g$y$ power spectrum, respectively. These spectra are plotted separately in the three panels 
of Fig.~\ref{fig:spectra}. In order to smooth out the power spectrum scatter across neighbouring multipoles, 
the data points show the bandpowers computed over a set of effective multipole bins $\elleff$, adopting the 
same binning scheme as in~\citet{Planck2016}. Overall, we consider $N_{\rm bands}=19$ multipole bins from 
a minimum $\elleff=10$ to a maximum $\elleff=1247.5$; the corresponding bandpowers are computed by averaging 
the spectrum values in each bin. In Table~\ref{tab:spectra} we report the values $\ell_{\rm eff}$ we consider, 
together with their multipole range limits and the bandpowers for $yy$, gg and g$y$. This binning procedure 
will also be applied to the theoretical prediction of the power spectra as described in Sec.~\ref{ssec:pseudocl}.
The spectra plotted in Fig.~\ref{fig:spectra} confirm that our data set allows measuring the auto-correlation 
of the galaxy density field and the Compton parameter maps. More importantly, a cross-correlation between 
the two is detected. It is convenient to quote the significance of these measurements, computed as
\begin{eqnarray}
	s = \left( C^{\rm T}\, \text{Cov}^{-1}\,C\right)^{1/2},  \label{eq:signif}
\end{eqnarray}
where $C$ represents any of the three power spectra and $\text{Cov}^{-1}$ is the inverse of its covariance 
matrix, which is required to account for possible correlations between multipole pairs; the superscript 
``T'' denotes transposition. The analysis of the covariance is described in details in Sec.~\ref{ssec:covmatr}; 
it is worth mentioning here that the diagonal terms of the covariance matrix quantify the uncertainties for the 
measured correlation at each multipole $\ell_{\rm eff}$, and are plotted as error bars in Fig.~\ref{fig:spectra}. 
The significance computed with Eq.~(\ref{eq:signif}) are $s=88$ for the gg spectrum, $s=150$ for the $yy$ spectrum 
and $s=21.8$ for the g$y$ spectrum. The latter value confirms that our measurement of the tSZ and galaxy 
density cross-correlation is robust. 

Finally, it is worth mentioning that our reconstruction of the $yy$ power spectrum is consistent with the results 
obtained in~\citet{Planck2016_sz}. To this aim, it is possible to compare the $yy$ amplitudes listed in Table~\ref{tab:spectra}
with the ones from Table~3 in~\citet{Bolliet18}. Marginal differences can be due to details in the data processing, such as 
the fact that we are using a non-apodized version of the mask, or the fact that \textit{Planck} considers
cross-correlations between data from the first and second halves of each pointing period, while we compute the auto-correlation 
from the full $y$ map. Also, the contribution of our 
non-Gaussian uncertainties (Sec.~\ref{ssec:covmatr}) implies our final error bars on the $yy$ power spectrum are larger than the 
ones shown in~\citet{Planck2016_sz}.

\begin{table*}
\centering
\caption{Summary of our power spectra measurements. By labelling $AB$ any of the possible combinations gg, g$y$ or $yy$, 
	we report for each effective multipole $\ell_{\rm eff}$ the power spectra value $C^{\rm AB}_{\ell}$, the Gaussian 
	error contribution $\sigma_{\rm G}(C^{\rm AB}_{\ell})$ and the non-Gaussian error contribution $\sigma_{\rm NG}(C^{\rm AB}_{\ell})$.}
\label{tab:spectra}	
\renewcommand{\arraystretch}{1.25}
\begin{tabular}{lccccccccc}
\hline 
  $\ell_{\rm eff}$ &$10^{16} C^{yy}_{\ell} $ &$  10^{17} \sigma_{\rm G}(C^{yy}_{\ell})$ & $10^{15} \sigma_{\rm NG}(C^{yy}_{\ell})$  &$10^{5} C^{\rm gg}_{\ell}$ &  $ 10^{6} \sigma_{\rm G}(C^{\rm gg}_{\ell})$ & $ 10^{5} \sigma_{\rm NG}(C^{\rm gg}_{\ell})$  & $ 10^{10} C^{\text{g}y}_{\ell}$  &$10^{11} \sigma_{\rm G}(C^{\text{g}y}_{\ell})$ & $ 10^{11} \sigma_{\rm NG}(C^{\text{g}y}_{\ell})$  \\
\hline
 10.0  &   3.91571  &  6.4590  &   6.10066  &   2.97111  &  7.92646  &  1.63821 &  1.46062 & 2.41424 &   1.32021  \\
 13.5  &   3.94016  &  4.9640  &   3.45057  &   1.80787  &  4.02548   &  1.08076 &  0.10616 & 1.12543 &   1.14705  \\
 18.0  &   3.41793  &  3.3880  &   1.86047  &   0.98915  &  1.39824   &  0.70758 &  0.27136 & 0.61249 &   0.96432   \\
 23.5  &   3.20770  &  2.4750  &   0.99994  &   0.69994  &  0.77641   &  0.46836 &  0.19546 & 0.37336 &   0.79265   \\
 30.5  &   1.84170  &  1.0460  &   0.52911  &   0.49552  &  0.40513   &  0.30773 &  0.12567 & 0.17285 &   0.63258   \\
 40.0  &   1.11424  &  0.5023  &   0.26011  &   0.34141  &  0.23748   &  0.19186 &  0.06377 & 0.08832 &   0.47585   \\
 52.5  &   0.79511  &  0.2730  &   0.13473  &   0.24317  &  0.13480   &  0.12306 &  0.06125 & 0.04923 &   0.35356   \\
 68.5  &   0.55403  &  0.1455  &   0.06832  &   0.18915  &  0.07900   &  0.07731 &  0.03724 & 0.02693 &   0.25163   \\
 89.5  &   0.44936  &  0.0903  &   0.03437  &   0.13320   & 0.04081   &  0.04802 &  0.02828 & 0.01556 &   0.17274   \\ 
 117.0 &   0.39211  &  0.0607  &   0.01717  &   0.10622  &  0.02576   &  0.02951 &  0.02624 & 0.01029 &   0.11489   \\
 152.5 &   0.32496  &  0.0390  &   0.00859  &   0.08248  &  0.01530   &  0.01802 &  0.02055 & 0.00625 &   0.07480   \\
 198.0 &   0.24493  &  0.0225  &   0.00431  &   0.05800  &  0.00789   &  0.01092 &  0.01408 & 0.00346 &   0.04801   \\ 
 257.5 &   0.20862  &  0.0147  &   0.00213  &   0.04301  &  0.00440   &  0.00648 &  0.01052 & 0.00210 &   0.03023   \\
 335.5 &   0.18025  &  0.0098  &   0.00121  &   0.03231  &  0.00247   &  0.00376 &  0.00860 & 0.00131 &   0.01868   \\
 436.5 &   0.15264  &  0.0064  &   0.00092  &   0.02550  & 0.00144   & 0.00214 &  0.00696 & 0.00082 &   0.01138   \\ 
 567.5 &   0.13614  &  0.0044  &   0.00055  &   0.02028	&  0.00083   &  0.00119 &  0.00499 & 0.00053 &   0.00680   \\
 738.5 &   0.12221  &  0.0030  &   0.00031  &   0.01520  &  0.00043   &  0.00065 &  0.00351 & 0.00033 &   0.00396   \\
 959.5 &   0.11573  &  0.0022  &   0.00019  &   0.01152  &  0.00022   &  0.00034 &  0.00270 & 0.00021 &   0.00224   \\ 
 1247.5&   0.12589  &  0.0020  &   0.00011  &   0.00898  &  0.00013   &  0.00018 &  0.00187 & 0.00017 &   0.00122   \\
\hline
\end{tabular}
\end{table*}

\subsection{Shot-noise correction}
The discrete nature of the WISE galaxy catalogue and its splitting into different pixels induces a shot-noise 
component in the overdensity map that can bias our computation of the gg power spectrum. In order to estimate 
the shot noise power spectrum, we follow the same procedure outlined in~\citet{Ando_2MASS} and~\citet{Makiya2018}. 
We randomly split the WISE catalog into two halves with a similar number of galaxies, and project them onto the 
sky generating two maps which we label $\delta_{\rm g,1}$ and $\delta_{\rm g,2}$. We then compute the half-sum 
(HS) and half difference (HD) maps as:
\begin{equation}
\text{HS} = \frac{1}{2} \left( \delta_{\rm g,1} + \delta_{\rm g,2} \right), \qquad \text{HD} = \frac{1}{2} \left( \delta_{\rm g,1} - \delta_{\rm g,2} \right)  \label{eq:HS-HD} .
\end{equation} 
By construction, the HS map contains both signal and noise, while in the HD map, the signal cancels out, 
leaving only the shot noise contribution. We then get to a noise-cleaned estimate of the gg power spectrum as
\begin{equation}
C^{\rm gg}_{\ell} = C^{\rm gg,HS}_{\ell} - C^{\rm gg,HD}_{\ell}.
\end{equation} 
The power spectrum plotted in the first panel of Fig.~\ref{fig:spectra} has already been shot noise-corrected. 
The correction is not applied to the cross-correlation of the galaxy overdensity field with the Compton map, 
as the shot noise in the former is uncorrelated from the noise affecting the latter. 

\subsection{The covariance matrix}
\label{ssec:covmatr}
The covariance matrix is required to complete the statistical characterisation of the measured power spectra, 
as well as to quantify their agreement with our theoretical models  (Sec.~\ref{ssec:likelihood}). We expect to 
observe a non-zero correlation between different multipoles, with increasing statistical weight as their separation 
decreases. In order to maximise the statistical information provided by our measurements, in this paper we combine 
the contribution of all of the observed spectra together. To this aim, we define a $3N_{\rm band}$-length vector 
$\mathbf{C}^{\rm tot}_{\ell}$ obtained by concatenating the three spectra gg, g$y$ and $yy$ as:
\begin{equation}
\label{eq:vector}
	\mathbf{C}^{\text{tot}}_{\ell}=\left\{C^{\rm gg}_{\ell}, C^{{\rm g}y}_{\ell}, C^{yy}_{\ell}\right\}.
\end{equation}
We need to derive a general matrix ${\rm Cov}^{\rm tot}$ that quantifies the covariance for the full vector 
$\mathbf{C}^{\text{tot}}_{\ell}$ (i.e the three spectra gg, g$y$ and $yy$ and the cross-correlation between 
different observed spectra). Applying the definition of covariance, it can be computed as:
\begin{eqnarray}
\label{eq:covfull}
	{\rm Cov}^{\rm tot} & =& \left\langle \mathbf{C}^{\text{tot}}_{\ell}{\mathbf{C}^{\text{tot}}_{\ell}}^{\rm T} \right\rangle  \nonumber \\
	& =& \left[
\begin{array}{ccc}
{\rm Cov}^{\rm gg,gg} & {\rm Cov}^{{\rm gg,g}y} & {\rm Cov}^{{\rm gg},yy} \\
\left({\rm Cov}^{{\rm gg,g}y} \right)^{\rm T} & {\rm Cov}^{{\rm g}y, {\rm g}y} & {\rm Cov}^{{\rm g}y, yy} \\
\left({\rm Cov}^{{\rm gg},yy} \right)^{\rm T} & \left({\rm Cov}^{{\rm g}y, yy} \right)^{\rm T} & {\rm Cov}^{yy,yy} \\
\end{array}
\right], 
\end{eqnarray}
where the brackets $\langle\rangle$ denote the statistical sample average. The last equality shows that the 
full covariance can be expressed as a set of six independent $N_{\rm band}\times N_{\rm band}$ covariance 
matrices of the form $\text{Cov}^{\rm AB,CD}(C^{\rm AB}_{\ell_1},C^{\rm CD}_{\ell_2})$, where again capital 
letters denote any one of $y$ or g. We recognise that the diagonal blocks, for which ${\rm AB}={\rm CD}$, are 
the covariance matrices for each of the correlations gg, g$y$ and $yy$, while the off-diagonal blocks quantify 
the ``cross'' covariance between different measured spectra. The square roots of the diagonal elements in the 
$N_{\rm band} \times N_{\rm band}$ matrices $\text{Cov}^{\rm gg,gg}$, $\text{Cov}^{\text{g}y,\text{g}y}$ and 
$\text{Cov}^{yy,yy}$ provide an estimate of the uncertainties associated with the corresponding power spectrum 
measurements, and are represented by the error bars in Fig.~\ref{fig:spectra}. By construction, these uncertainties 
also carry the contribution from any foreground residual contamination in the maps.

The covariance matrix needs to be evaluated analytically. Under the assumption that the thermal SZ fluctuations 
are purely Gaussian, the covariance matrix would be diagonal with no correlation between different multipoles. 
It could be evaluated from the knowledge of the measured spectra and the available sky fraction. However, works 
on hydrodynamical simulations proved that SZ fluctuations can indeed be non-Gaussian~\citep{Seljak_Burwell_2001,Zhang_pen_Wang_2002}. 
We then write the covariance as the sum of a Gaussian and a non-Gaussian term following~\citet{Makiya2018}: 
\begin{eqnarray}
\label{eq:Theoretical_covariance}
{\rm Cov}^{\rm AB,CD}(C^{\rm AB}_{\ell_1},C^{\rm CD}_{\ell_2} ) &=&  {\rm Cov}^{\rm G} (C^{\rm AB}_{\ell_1},C^{\rm CD}_{\ell_2} )  \nonumber \\
&+& {\rm Cov}^{\rm NG} (C^{\rm AB}_{\ell_1},C^{\rm CD}_{\ell_2} ).
\end{eqnarray} 
We write the Gaussian component as
\begin{eqnarray}
\label{eq:Gaussian_Term}
&&{\rm Cov}^{\rm G}(C^{\rm AB}_{\ell_1},C^{\rm CD}_{\ell_2} ) \nonumber \\ && =\frac{[\hat{C}^{\rm AC}_{\ell_1}\,\, \hat{C}^{\rm BD}_{\ell_2} + \hat{C}^{\rm AD}_{\ell_1}\,\, \hat{C}^{\rm BC}_{\ell_2}]}{f^{\rm ABCD}_{\rm sky}(2 \ell_1 + 1)\Delta \ell_1} \delta_{{\ell_1}{\ell_2}},
\end{eqnarray} 
where $f^{\rm ABCD}_{\rm sky}$ is the observed sky fraction. The spectra in the square brackets (i.e., 
$\hat{C}^{\rm AC}_{\ell}, \hat{C}^{\rm AD}_{\ell}, \hat{C}^{\rm BD}_{\ell}$ and $\hat{C}^{\rm BC}_{\ell}$ ) are 
the observed power spectra which include the contribution from noise and foregrounds, $\delta_{{\ell_1}{\ell_2}}$ 
is the Kronecker delta, and $\Delta \ell$ gives the discrete difference between multipole bins. We write the 
non-Gaussian term of the covariance matrix as
\begin{eqnarray}
\label{eq:non-Gaussian_Term}
	{\rm Cov}^{\rm NG}(C^{\rm AB}_{\ell_1},C^{\rm CD}_{\ell_2} )  = \frac{T^{\rm ABCD}_{{{\ell_1}{\ell_2}}}}{4 \pi \,f^{\rm ABCD}_{\rm sky}}.
\end{eqnarray} 
The angular trispectrum is given by (see also \citealt{Makiya2018} and \citealt{Bolliet18}):
\begin{eqnarray}
\label{eq:trispectrum_Term}
T^{\rm ABCD}_{{{\ell_1}{\ell_2}}} & = & \int \textrm{d}z \frac{c \chi^2(z)}{H(z)} \int \textrm{d}M \frac{{\rm d}n}{{\rm d}M}(M,z)\, A_{\ell_1}(M,z) \nonumber \\
& \times & B_{\ell_1}(M,z) C_{\ell_2}(M,z) D_{\ell_2}(M,z), 
\end{eqnarray}
where $\chi(z)$ is the comoving distance, $H(z)$ is the Hubble parameter and ${\rm d}n/{\rm d}M$ is the 
halo mass function. The sky fraction in Eq.~(\ref{eq:non-Gaussian_Term}) depends on the chosen mask. As we used different masks on the WISE 
and tSZ maps, $f^{\rm ABCD}_{\rm sky}$ is determined by the combinations of the two. The sky fraction is equal 
to $f^{yy}_{\rm sky} = 0.59$ and $f^{\rm gg}_{\rm sky} = 0.40$ for the $yy$- and gg-auto correlation, respectively. 
We determined the sky fraction for the cross-correlation as 
$f^{\rm gg,yy}_{\rm sky}=\sqrt{f^{\rm gg}_{\rm sky}f^{\rm yy}_{\rm sky}} \equiv 0.486$~\citep{Makiya2018}.
In Table~\ref{tab:spectra} we report the diagonal values of the Gaussian and non-Gaussian covariance matrices 
obtained for $yy$, gg and g$y$.  

\begin{figure*}
	\centering
	\includegraphics[width=17cm]{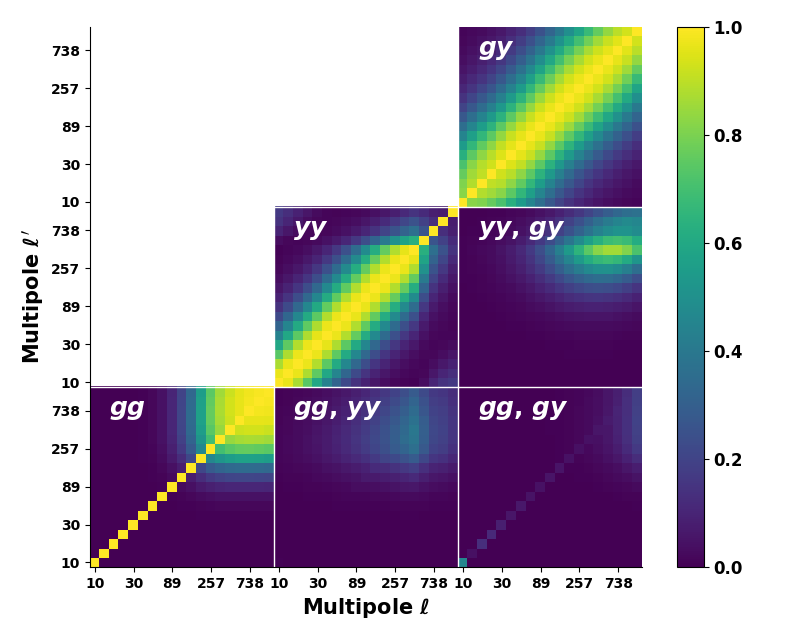}
	\caption{The six independent blocks of the full correlation matrix defined in 
	         Eq.~(\ref{eq:correlation}). The diagonal blocks show the standard 
		 correlations for the gg, $yy$ and g$y$ spectra; the off-diagonal blocks
		 show the additional cross-correlations between different spectra. 
		 The correlation values are shown for pairs of the effective bandpower 
		 multipoles defined in Table~\ref{tab:spectra}.}
	\label{fig:correlation}
\end{figure*}

Before closing this section we compute the correlation coefficient matrix, which can be obtained from the full 
covariance by normalising it to the diagonal values of the associated $N_{\rm band}\times N_{\rm band}$ covariance 
matrices:
\begin{eqnarray}
	\label{eq:correlation}
	&&{\rm Corr}^{\rm AB,CD}(\ell,\ell') \nonumber \\
	&& = \dfrac{{\rm Cov}^{\rm AB,CD}(\ell,\ell')}{\sqrt{{\rm Cov}^{\rm AB, AB}(\ell,\ell)}\sqrt{{\rm Cov}^{\rm CD,CD}(\ell',\ell')}}.
\end{eqnarray}
Notice that, according to this definition, only the diagonal elements of the diagonal blocks of the covariance 
matrix from Eq.~(\ref{eq:covfull}) are equal to unity, whereas the diagonal elements in the non-diagonal blocks 
are not normalised to one. The resulting correlation matrix, made by its six independent building blocks, is 
plotted in Fig.~\ref{fig:correlation}, binned over the effective multipoles $\ell_{\rm eff}$. The off-diagonal 
elements are negligible in each block due to the increasing multipole separation and the binning over effective 
multipoles. On the contrary, the non-zero diagonals inside the off-diagonal blocks reflect the existing correlation 
between $C_{\ell}^{\rm gg}$, $C_{\ell}^{\text{g}y}$ and $C_{\ell}^{yy}$, which is non-negligible. Such correlation 
is indeed captured by the full covariance matrix computed with Eq.~(\ref{eq:Theoretical_covariance}), which is the 
one we employ to conduct the parameter estimation in Sec.~\ref{sec:parest}.


\section{Theoretical modeling}
\label{sec:modelling}

We detail in this section the formalism we employ to theoretically predict the observed auto-correlation power 
spectra $\Ctsz$ and $\Cgal$, and the cross-correlation power spectrum $\Ccross$ theoretically. The following describes 
how the mass bias parameter we are interested in enters this prediction, together with a set of nuisance parameters 
whose correlation will be explored in Sec.~\ref{sec:parest}.

\subsection{Halo model}
\label{ssec:halomodel}

Our theoretical framework for the angular power spectrum calculation is based on the halo model, which is an 
established approach to the problem of predicting cross-correlations under the assumption that all galaxies live 
in haloes~\citep{Kita_and_komatsu,Ken-Osato2019}. We label $C^{\rm AB}_{\ell}$ the generic cross-correlation 
power spectrum between observables ${\rm A}$ and ${\rm B}$. According to the halo model, it can be decomposed 
into the contribution of a one-halo (intra-halo) term $C^{\rm AB,\text{1h}}_{\ell}$ and of a two-halo (inter-halo) 
term $C^{\rm AB,\text{2h}}_{\ell}$, as:
\begin{equation}
\label{eq:halomodel}
	C^{\rm AB}_{\ell} = C^{\rm AB,\text{1h}}_{\ell} + C^{\rm AB,\text{2h}}_{\ell}.
\end{equation}

The one-halo term quantifies the integrated contribution of all the observable haloes considered individually, 
and can be expressed as:  
\begin{eqnarray} 
\label{eq:1halo}
	C^{\rm AB,\text{1h}}_{\ell} & =& \int_{z_{\rm min}}^{z_{\rm max}} \textrm{d}z \frac{c \chi^2(z)}{H(z)}   \int^{M_{\rm max}}_{M_{\rm min}} \textrm{d}M \frac{{\rm d}n}{{\rm d}M}(M,z)\,   \nonumber \\
	&  \times & A_{\ell}(M,z) B_{\ell}(M,z),  
\end{eqnarray}
where $c \chi^2(z)/H(z) = {\rm d}^2V/({\rm d}z {\rm d}\Omega)$ is the comoving volume per unit redshift and solid angle, and $A_{\ell}$ and $B_{\ell}$ are the spherical Fourier transforms of 
the corresponding generic observables on the sky. The integral endpoints $z_{\rm min}$, $z_{\rm max}$ and $M_{\min}$, 
$M_{\max}$ are to be chosen depending on the redshift and mass spans of the targeted observables. 

The two-halo term quantifies the effect of inter-halo correlations, and can be written as:
\begin{align}
	\label{eq:2halo}
	&C^{\rm AB,\text{2h}}_{\ell} = \int_{z_{\rm min}}^{z_{\rm max}} \textrm{d}z \frac{c \chi^2(z)}{H(z)} P^{\rm{lin}}_{\rm{m}}\left(k = \frac{\ell +1/2}{\chi(z)},z\right) \nonumber \\
&\times \left[\int^{M_{\rm max}}_{M_{\rm min}}{\rm d}M\frac{{\rm d}n}{{\rm d}M}(M,z)\,b(M,z)A_{\ell}(M,z)\right] \nonumber \\
&\times   \left[\int^{M_{\rm max}}_{M_{\rm min}}{\rm d}M\frac{{\rm d}n}{{\rm d}M}(M,z)\,b(M,z)B_{\ell}(M,z)\right],  
\end{align} 
where $P^{\rm{lin}}_{\rm{m}}(k,z)$ is the linear matter power spectrum and $b(M,z)$ is the halo bias. 

In our implementation we will use the mass function parametrisation from~\citet{Tinker2008}, the halo 
bias parametrisation from~\citet{Tinker2010} and compute the linear matter power spectrum using 
{\sc camb}~\citep{Lewis2000}. As per the integration extrema, we set $z_{\min}=10^{-3}$, $z_{\max}=5$ when 
computing the $yy$ auto-correlation, as this interval safely includes all contributions from galaxy clusters 
(we do not set $z_{\min}=0$ to avoid divergences in the computation of angular sizes). For the g$y$ cross-correlation 
and the gg auto-correlation, instead, we set $z_{\min}=3\times10^{-2}$ and $z_{\max}=1$, as this the redshift 
range spanned by WISE galaxies (see Fig.~\ref{fig:dndz-fit}). Regarding the mass, we set a lower limit of 
$10^{11}\,h^{-1}\text{M}_{\odot}$, below which the ICM pressure becomes negligibly low and an upper limit 
of $10^{16}\,h^{-1}\text{M}_{\odot}$, after which the mass function severely cuts off the halo abundance. 
We checked that the final predictions do not vary appreciably if changing the mass limits by a few per cent in $\log_{10}(M)$.

The remaining quantities to be determined are the Fourier transforms $A_{\ell}(M,z)$ and $B_{\ell}(M,z)$ 
for the generic observables. We detail in the rest of this section their evaluation for the Compton parameter 
and the galaxy density field. Before, it is worth reminding that the mass function is parametrised in terms of 
the overdensity mass $M_{\rm 200,m}$~\citep{Tinker2008}, i.e. the mass enclosing a radius whose mean density 
equals 200 times the matter cosmic density at that redshift. The cluster pressure profile, which is used to 
compute the Compton parameter, is instead typically expressed as a function of $M_{\rm 500,c}$, namely the mass 
enclosing a radius $R_{500}$ whose mean density equals 500 times the critical density of the Universe $\rho_{\rm crit}$:
\begin{equation}
\label{eq:m500}
M_{\rm 500,c} = \frac{4}{3} \pi \left[500 \rho_{\rm{crit}}(z)\right]R^3_{500}.
\end{equation}
The critical density, in turn, can be expressed as $\rho_{\rm crit}(z) = 2.77 \times 10^{11} E^2(z)\,h^2 \,\rm{M}_{\odot} \rm{Mpc}^{-3}$,
where $E(z)=H(z)/H_{0}=\sqrt{\Omega_{\rm m}(1+z)^{3}+\Omega_{\Lambda}}$ with $\Omega_{\rm m}$ and $\Omega_{\Lambda}$ the 
matter and dark energy density parameters, respectively. Finally, the galaxy density field is usually modelled 
via the halo virial mass $M_{\rm vir}$, defined as: 
\begin{equation}
\label{eq:mvir}	
M_{\rm{vir}} = \frac{4}{3} \pi \left[\Delta_{\rm vir}(z) \rho_{\rm{crit}}(z)\right]R^3_{\rm{vir}} ,
\end{equation} 
where $R_{\rm{vir}}$ is the virial radius and the overdensity $\Delta_{\rm vir}$ is parametrised in~\citet{bryan98} 
as (see also eqs.~D2, D9, D10 in~\citealt{komatsu11})
\begin{equation}
\label{eq:deltavir}
	\Delta_{\rm vir} = 18\pi^2 + 82\left[\Omega(z)-1\right] -39\left[\Omega(z)-1\right]^2,
\end{equation}
with
\begin{equation}
\Omega(z) = \Omega_{\rm m}\frac{(1+z)^2}{E^2(z)}.
\end{equation}

In our implementation of the halo model formalism, we choose to set $M_{\rm 200,m}$ as our independent variable, 
and the aforementioned integration limits are quoted for this mass definition. When computing the Fourier transform 
for the Compton parameter or the galaxy density field, the value is properly converted into $M_{\rm 500,c}$ and 
$M_{\rm vir}$, respectively. This conversion is performed employing the {\tt Python} \textsc{COLOSSUS} 
package\footnote{\url{https://bdiemer.bitbucket.io/colossus/index.html}.}~\citep{Colossus2018} 
assuming a Navarro–Frenk–White~\citep[NFW,][]{navarro96} profile and the mass-concentration relation from~\citet{Duffy2008}.


\subsection{Fourier space Compton-$y$ parameter} 
\label{ssec:yfourier}

The Compton parameter defined in Eq.~(\ref{eq:compton}) is proportional to the electron pressure
$P_{\rm e} = k_{\rm B} n_{\rm e} T_{\rm e}$ integrated along the line of sight. The effect is 
particularly relevant in the direction of galaxy clusters; assuming that galaxies reside in 
virialised dark matter haloes, the $y$ definition allows to define an effective halo
2-dimensional Compton parameter profile projected on the sky. For a generic halo of mass
$M_{\rm 500}$ (to simplify the notation we will drop the subscript ``c'' hereafter) at redshift 
$z$, the latter is a function of the 3-dimensional electron pressure profile 
$P_{\rm e}(r;M_{\rm 500},z)$, with $r$ the comoving radial separation from the halo center. 

The associated 2-dimensional SZ Fourier transform for a single halo can then be computed 
as~\citep{Planck2014_tSZ,Planck2016_sz}:
\begin{align}
y_{\ell}(M_{500},z) &=  \frac{4\pi R_{\rm 500}}{\ell^2_{\rm 500}} \frac{\sigma_{\rm T}}{m_{\rm e} c^2} \nonumber \\
	&\times \int {\rm d}x\, x^2 \frac{\sin(\ell x/\ell_{\rm 500})}{\ell x/\ell_{\rm 500}} P_{\rm e}(x;M_{500},z),
\end{align} 
where we introduced the scaled radial separation $x=a\,r/R_{500}$ (with $a$ the scale factor
at the halo redshift), and $\ell_{\rm 500} =a \chi/R_{\rm 500}$. We evaluate the integral between the 
limits $x_{\rm min}=0$ and $x_{\rm max}=6$ as the physical scale $5\,R_{500}$ is usually considered to 
mark the outer boundary of a galaxy cluster. We adopt the electron pressure profile parametrisation 
derived in~\citet{Arnaud2010}:
\begin{eqnarray}
\label{eq:pressprof}
&& P_{\rm e} \left(x;M_{500},z \right)  =  1.65\,\, h^{2}_{70}\,\, E^{8/3}(z)  \nonumber \\
&& \times   \left[ \frac{(1-b_{\rm H})M_{500}}{3 \times 10^{14}h^{-1}_{70} M_\odot} \right]^{2/3 + \alpha_{\rm p}} \mathbb{P}(x) \, {\rm [eV\,cm^{-3}]},  
\end{eqnarray} 
where $h_{70} = h/0.7$, $\alpha_{\rm p}\simeq 0.12$ represents the departure from the standard self-similar 
solution and $\mathbb{P}(x)$ is the ``universal'' pressure profile (UPP). The latter is parametrised as a generalised 
Navarro-Frenk-White profile~\citep{nagai07}:
\begin{equation}
	\label{eq:upp}
\mathbb{P}(x) = \frac{P_0}{(c_{500}x)^\gamma \left [1 + (c_{500}x)^\alpha   \right ]^{(\beta - \gamma)/\alpha}},
\end{equation} 
which we compute using the parameter values from~\citet{Planck2013}, 
$\left\{P_0,c_{500},\alpha,\beta,\gamma\right\}=\left\{6.41,1.81,1.33,4.13,0.31\right\}$, fitted over a set of 62 
nearby massive clusters observed by \textit{Planck}. Finally, Eq.~(\ref{eq:pressprof}) shows that the pressure 
depends on the effective tSZ cluster mass $M^{\rm tSZ}_{500}$ already introduced in Eq.~(\ref{eq:biasdef}).
The departure from the assumption of hydrostatic equilibrium in the ICM, together with other model systematics, is 
quantified by the bias parameter $B$ or by its equivalent $b_{\rm H}$. We want to stress that we also account for the 
hydrostatic mass bias when computing $R_{500}$, so that our definition has been rescaled as $R_{500}(1-b_{\rm H})^{1/3}$.

As anticipated in Sec.~\ref{sec:intro}, providing an independent estimate of $b_{\rm H}$ is a major goal of our paper. 
We find it more convenient to fit for the bias expressed as $B$, which will be the independent parameter in the analysis 
described in Sec.~\ref{ssec:likelihood}. We will also report the corresponding constraints on the quantity $1-b_{\rm H}$ in 
Table~\ref{tab:Bias_comparison}.


\subsection{Galaxy density field Fourier transform} 
\label{ssec:gfourier}
The projected galaxy density field measured at a generic direction $\hat{\mathbf{n}}$ on the sky can generally 
be expressed by integrating the matter overdensity $\delta_{\rm m}$ over the comoving distance 
as~\citep{Ferraro2015,hill16,ferraro16}:
\begin{equation}
\label{eq:densfield}
\delta_{\rm g}(\hat{\mathbf{n}}) = \int\,\text{d}\chi \, W_{\rm g}(\chi)\, \delta_{\rm m}(\chi \hat{\mathbf{n}}),
\end{equation} 
where the kernel function $W_{\rm g}(\chi)=b_{\rm g}p_{\rm s}(\chi)$ is the product of the galaxy bias 
$b_{\rm g}$ and the source distribution function $p_{\rm s}(\chi)$. The latter quantifies the probability 
of detecting a source in the interval $[\chi,\chi+\text{d}\chi]$, and has to satisfy the normalisation condition:
\begin{equation}
	\int^{\infty}_{0} \text{d}\chi \, p_{\rm s}(\chi)=1.
\end{equation}
The source function can be more conveniently expressed as a function of redshift via the variable change 
\begin{equation}
	p_{\rm s}(z)=p_{\rm s}(\chi) \frac{\text{d}\chi}{\text{d}z} =p_{\rm s}(\chi)\frac{c}{H(z)}
\end{equation}
(we will continue to call the redshift distribution $p_{\rm s}$ in a slight abuse of notation). For our WISE 
catalogue, the source distribution as a function of redshift is plotted in Fig.~\ref{fig:dndz-fit}. 

The galaxy bias $b_{\rm g}$ is an unconstrained quantity in our model. To a first approximation, it could be 
factorised out of the integral in Eq.~(\ref{eq:densfield}) as a mean value for our data set, assuming it is 
independent of redshift; such approach was adopted for example in \citet{ferraro16} and~\citet{hill16}. Given 
the wide range of $z$ values spanned by the WISE catalogue, it is more meaningful to explore a possible redshift 
dependence of the halo bias. Furthermore, we notice that the high-$\ell$ points have the smallest error bars and 
consequently a higher statistical weight in the parameter estimation described in Sec.~\ref{sec:parest}. As those 
points lie in the range where the one-halo term is dominant, the latter is expected to be driving the fit in 
determining the most likely value for the galaxy bias. In order to break this coupling between small and large 
scales, and increase the statistical weight of the two-halo term in the fit for $b_{\rm g}$, we also include an 
explicit bias dependence on the multipole. As a result, we parametrise the galaxy bias as:
\begin{equation}
	\label{eq:bz}
	b_{\rm g}(z,\ell) = b_{\rm g}^0 \, (1 + z)^{\alpha} \left(\frac{\ell}{\ell_0}\right)^{\beta},
\end{equation}
letting the normalisation $b_{\rm g}^0$ at $z=0$, $\ell=\ell_0$ and the scaling power indices $\alpha$ and 
$\beta$ be free parameters in our model. The pivot scale $\ell_0=117$ is computed as the median of the 
available multipole range, and it roughly corresponds to the scale at which the one- and two-halo terms 
have comparable amplitudes. The parametrisation in Eq.~(\ref{eq:bz}) is a generalisation of the redshift 
dependence for $b_{\rm g}$ explored in~\citet{Ferraro2015}.

For a generic halo of mass $M$ at redshift $z$, let $\rho(ar;M,z)$ be the matter density at a radial separation 
$ar$ from its center ($r$ being the comoving separation and $a$ the scale factor), and $\rho_{\rm m}(z)$ the mean 
matter density at the same redshift. The associated matter overdensity halo profile is then defined by the ratio: 
\begin{equation}
	\label{eq:deltam}
	\delta_{\rm m}(ar;M,z) = \frac{\rho(ar;M,z)}{\rho_{\rm m}(z)} - 1.
\end{equation}
The 2-dimensional Fourier transform for the matter overdensity defined above can be computed as:
\begin{eqnarray}
	\label{eq:deltam2d}
\delta_{\rm m,2D}(\ell;M,z) &=&  \int^{\infty}_{0} \text{d}r \,(4\pi r^{2}) \nonumber \\
&\times & \left(\frac{\sin(\ell r/\chi)}{\ell r/\chi}\right)\delta_{\rm m}(ar;M,z).
\end{eqnarray}
In order to Fourier transform the galaxy overdensity, we have to include the kernel function $W_{\rm g}$ 
introduced in Eq.~(\ref{eq:densfield}), as: 
\begin{equation}
	\label{eq:gell}
 g_{\ell}(M,z)=\frac{W_{\rm g}(z)}{\chi^{2}(z)}\delta_{\rm m,2D}\left(\frac{\ell}{\chi(z)}; M,z \right).
\end{equation}

The computation of the Fourier transform in Eq.~(\ref{eq:gell}) requires the choice of a 
functional form for the matter density halo profile $\rho(ar;M,z)$, for which we shall assume 
again an NFW parametrisation as:
\begin{equation}
	\label{eq:nfw}
\rho(R;M,z)=\frac{\rho_{0}}{\left(R/r_{\rm s}\right)\left(1+R/r_{\rm s} \right)^{2}}, 
\end{equation}
where we denote by $R$ the physical radial separation from the halo centre. The NFW profile is 
governed by two parameters, the normalisation density $\rho_0$ and the scale radius $r_{\rm s}$. 
The normalisation density can be computed by imposing that the volume integral of the halo density 
within its radius yields the total halo mass. As in this context we are working with virial 
quantities, the mass normalisation condition reads:
\begin{equation}
	\label{eq:normalisation}
	4\pi \int^{R_{\rm vir}}_{0} \text{d}R \, R^{2}\,\rho(R)=M_{\rm vir}. 
\end{equation}
By introducing the halo concentration parameter as the ratio between the virial and the scale 
radius, $c_{\rm vir}=R_{\rm vir}/r_{\rm s}$, the integral in~Eq.~(\ref{eq:normalisation}) can be 
carried out to yield the normalisation density:
\begin{equation}
	\label{eq:rho0}
	\rho_{0}=\frac{M_{\rm vir}}{4\pi r^{3}_{\rm s}m(c_{\rm vir})}, 
\end{equation}
where the function $m(x)$ is defined as $m(x)=\ln(1+x)-x/(1+x)$~\citep{Cooray02}. We shall use the parametrisation 
from~\citet{Duffy2008} to compute the concentration parameter of a generic halo of mass $M_{\rm vir}$ 
at redshift $z$:
\begin{equation}
	\label{eq:def-c}
	c_{\rm vir} =   \frac{5.72}{(1+z)^{0.71}} \left( \frac{M_{\rm vir}}{10^{14}h^{-1}\rm{\text{M}_{\odot}}} \right)^{-0.081}. 
\end{equation}
 
The normalised NFW profile can then be plugged back into Eqs.~(\ref{eq:deltam}) and~(\ref{eq:deltam2d}) 
to compute the 2-dimensional Fourier transform of the matter overdensity field. By defining the scaled 
radial separation as $x=ar/r_{\rm s}$, and $\ell_{\rm s}=a\chi/r_{\rm s}$, we obtain the expression:
\begin{eqnarray}
	\label{eq:delta-3d}
&& \delta_{\rm m, 2D}(\ell;M_{\rm vir},z)  =  4\pi \left(\frac{r_{\rm s}}{a} \right)^{3}   \nonumber \\
    && \times  	\int^{\infty}_{0} \text{d}x \,x^{2}\,\frac{\sin{(\ell x/\ell_{\rm s})}}{(\ell x/\ell_{\rm s})} \, \frac{\rho(x r_{\rm s}; M_{\rm vir},z)}{\rho_{\rm m}(z)} \nonumber \\
	&& = 
	\frac{4\pi r^{3}_{\rm s} }{\rho_{\rm crit}(0)\Omega_{\rm m} \tilde{\ell}} \nonumber \\
	&& \times \int^{\infty}_{0} \text{d}x \,x \sin\left(\tilde{\ell} x \right)\rho\left(x r_{\rm s} | M_{\rm vir},z \right), 
\end{eqnarray}
where $\tilde{\ell}\equiv\ell/\ell_{\rm s}$ and in the last step we made the matter density redshift 
dependence explicit, $\rho_{\rm m}(z) = \rho_{\rm crit}(0)\Omega_{\rm m}a^{-3}$ ($\rho_{\rm crit}(0)$ is the critical density at redshift $0$). By substituting the expression 
for the NFW profile from Eq.~(\ref{eq:nfw}) and its normalisation from Eq.~(\ref{eq:normalisation}), we obtain
\begin{eqnarray}
&&\delta_{\rm m,2D}(\ell; M_{\rm vir},z) = \frac{M_{\rm vir}}{\rho_{\rm crit}(0)\Omega_{\rm m}} \nonumber \\
&& \times \frac{1}{[\ln(1+c_{\rm vir})-c_{\rm vir}/(1+c_{\rm vir})]} \nonumber \\
&& \times  \left[\frac{\pi}{2}\sin{\tilde{\ell}}-\left(\cos{\tilde{\ell}}\,\text{Ci}\left(\tilde{\ell} \right) + \sin{\tilde{\ell}}\,{\rm Si}\left(\tilde{\ell}\right) \right) \right], 
\end{eqnarray} 
where we defined for convenience the sine and cosine integral functions:
\begin{equation}
	\text{Si}(x) = \int^x_0 \text{d}t \,\frac{\sin{t}}{t},  \qquad  \text{Ci}(x) = - \int^{\infty}_x \text{d}t \,\frac{\cos{t}}{t}.
\end{equation} 
The final expression for the Fourier transform of the galaxy overdensity field $g_{\ell}$ is therefore:
\begin{eqnarray}
	\label{eq:gell-M-z}	
	&&g_{\ell}(M_{\rm vir},z) = \frac{b_{\rm g}(z,\ell)p_{\rm s}(z) H(z)}{c \,\chi^{2}(z)} \frac{M_{\rm vir}}{\rho_{\rm crit}(0)\Omega_{\rm m}}    \nonumber \\
	&&  \times   \frac{(\pi\sin{\tilde{\ell}})/2-\left[\cos{\tilde{\ell}}\,\text{Ci}\left(\tilde{\ell} \right) + \sin{\tilde{\ell}}\,{\rm Si}\left(\tilde{\ell}\right) \right]}{\ln(1+c_{\rm vir})-c_{\rm vir}/(1+c_{\rm vir})}. 
\end{eqnarray}
The result in Eq.~(\ref{eq:gell-M-z}) can then be plugged into Eqs.~(\ref{eq:1halo}) and~(\ref{eq:2halo}) 
to compute either the auto-correlation power spectrum for the galaxy density field, or its cross-correlation 
power spectrum with the Compton parameter map. 

\begin{figure}
	\centering
	\includegraphics[width=8.5cm]{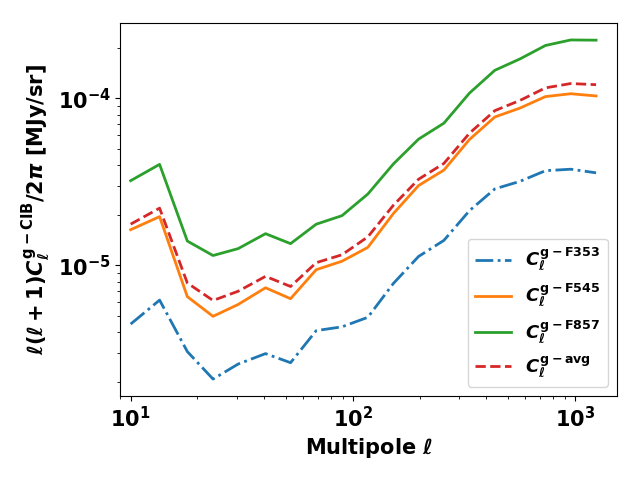}
	\caption{Cross-correlation between the WISE galaxy overdensity map and each of the three \textit{Planck} CIB 
		maps at different frequencies. Because the spectra show a similar multipole 
		dependence, we can take their average (red dashed line) to estimate the effect of 
		CIB contaminations in the g$y$ cross-correlation, as it is made explicit in Eq.~(\ref{eq:cibcl}).}
	\label{fig:cib}		
\end{figure}


\subsection{modeling the foreground contribution}
\label{ssec:foregrounds}

As anticipated in Sec.~\ref{ssec:planck}, the Compton parameter map is affected by residual foreground 
contaminations that are not completely suppressed in the component separation analysis. These contaminations 
will bias the tSZ power spectrum measured from the map in Sec.~\ref{sec:analysis}, so that the pure tSZ 
power spectrum model described in Sec.~\ref{ssec:yfourier} is no longer representative of the observed 
auto-correlation. We shall model the real $yy$ auto-correlation spectrum, which hereafter is labelled 
$C^{yy}_{\ell}$, as the halo-model predicted tSZ-only spectrum, $C^{\rm tSZ}_{\ell}$, plus a set of foreground terms as:
\begin{eqnarray}
\label{eq:clforegrounds}
C^{yy}_{\ell} &=&  \,C^{\rm tSZ}_{\ell} + A_{\rm CIB}\,C^{\rm CIB}_{\ell}+A_{\rm IR}\,C^{\rm IR}_{\ell}\nonumber \\
&+& A_{\rm Rad}\,C^{\rm Rad}_{\ell}+A_{\rm CN}\,C^{\rm CN}_{\ell}.
\end{eqnarray} 
The relation above considers the contribution from the clustered cosmic infrared background (CIB), infrared 
sources (IR), radio sources (Rad) and instrumental correlation noise (CN). The study of these foreground 
contributions to the \textit{Planck} tSZ map has already been tackled in previous works, and we will employ 
for our analysis their values tabulated in~\citet{Planck2016_sz}. Although such templates define the contaminants 
dependence at different angular scales, we let their actual contribution to the $yy$ power spectrum be controlled 
by the set of amplitude parameters $A_{\rm CIB}$, $A_{\rm IR}$ and $A_{\rm Rad}$, which are not constrained 
\textit{a priori} and shall be fitted against our observables. We only fix the value for $A_{\rm CN}=0.903$, 
as this value is required to reproduce the $yy$ spectrum at the highest multipole $\ell=2742$, where instrumental 
noise is the dominant contribution~\citep{Bolliet18,Makiya2018}.

As WISE data probe the galaxy overdensity field up to $z\sim 0.8$, the cross correlation with the Compton map may 
also be affected by CIB contaminations, as the latter is indeed a relevant foreground at 
$z \sim 1$~\citep{Makiya2018}. Our theoretical modeling should therefore incorporate this possible contribution 
in the prediction of the g$y$ cross-correlation. The \textit{Planck} Collaboration delivered three maps of CIB 
anisotropies outside the Galactic plane region at the frequencies 353, 545 and 857 GHz~\citep{planck_cib}. To assess the CIB contamination 
level in our power spectra, we compute the cross-correlation of each of these maps (downgraded to a 10' resolution) 
with the WISE galaxy overdensity map. The resulting spectra are plotted in Fig.~\ref{fig:cib}; the decrease in power at 
$\ell \sim 1000$ is due to the beam smoothing, while there is no straightforward interpretation for the observed peak at low-$\ell$. 
Using a different CIB map affects the amplitude of 
the resulting correlation, but does not result in an appreciable change in the spectrum shape. Therefore, we can 
consider the average of these three spectra, $C_{\ell}^{{\rm g }-{\rm CIB}_{\rm avg}}$, also plotted in 
Fig.~\ref{fig:cib}, to be representative of the CIB contamination dependence on $\ell$. The CIB contamination is 
then included in our modeling of the cross-correlation between WISE and \textit{Planck} data as:
\begin{equation}
\label{eq:cibcl}
	C^{\text{g}y}_{\ell} =  C^{{\rm g} - {\rm tSZ}}_{\ell} + B_{\rm CIB}C^{{\rm g }-{\rm CIB}_{\rm avg}}_{\ell},
\end{equation}
where $ C^{{\rm g}-{\rm tSZ}}_{\ell}$ is the cross-power spectrum between the Compton map and the galaxy overdensity 
map computed using the halo model (Eq.~(\ref{eq:halomodel})) and $B_{\rm CIB}$ is a free parameter which gauges the 
actual CIB contamination at the power spectrum level\footnote{With this formalism we are making the implicit assumption 
that the spectral dependence of the cross-correlation between WISE and the CIB maps is representative of the 
cross-correlation between WISE and the CIB residuals in the $y$ map. See similar treatment in~\citet{Alonso2018},~\citet{Yan2019} and~\citet{Yan2021}.}. $B_{\rm CIB}$ is then different from the 
parameter $A_{\rm CIB}$ which controls the amplitude of the CIB contamination in the $yy$ auto-correlation, as it 
is not dimensionless. Since the CIB maps report the foreground specific intensity in units $\text{MJy}\,\text{sr}^{-1}$, 
while the galaxy overdensity map is dimensionless, we expect the amplitude coefficient $B_{\rm CIB}$ to have dimensions ${\rm sr}\,{\rm MJy}^{-1}$. Its actual value has to be fitted against our measurements, as it is
described in Sec.~\ref{ssec:likelihood}.


\begin{figure*}
	\centering
	\includegraphics[width=18cm]{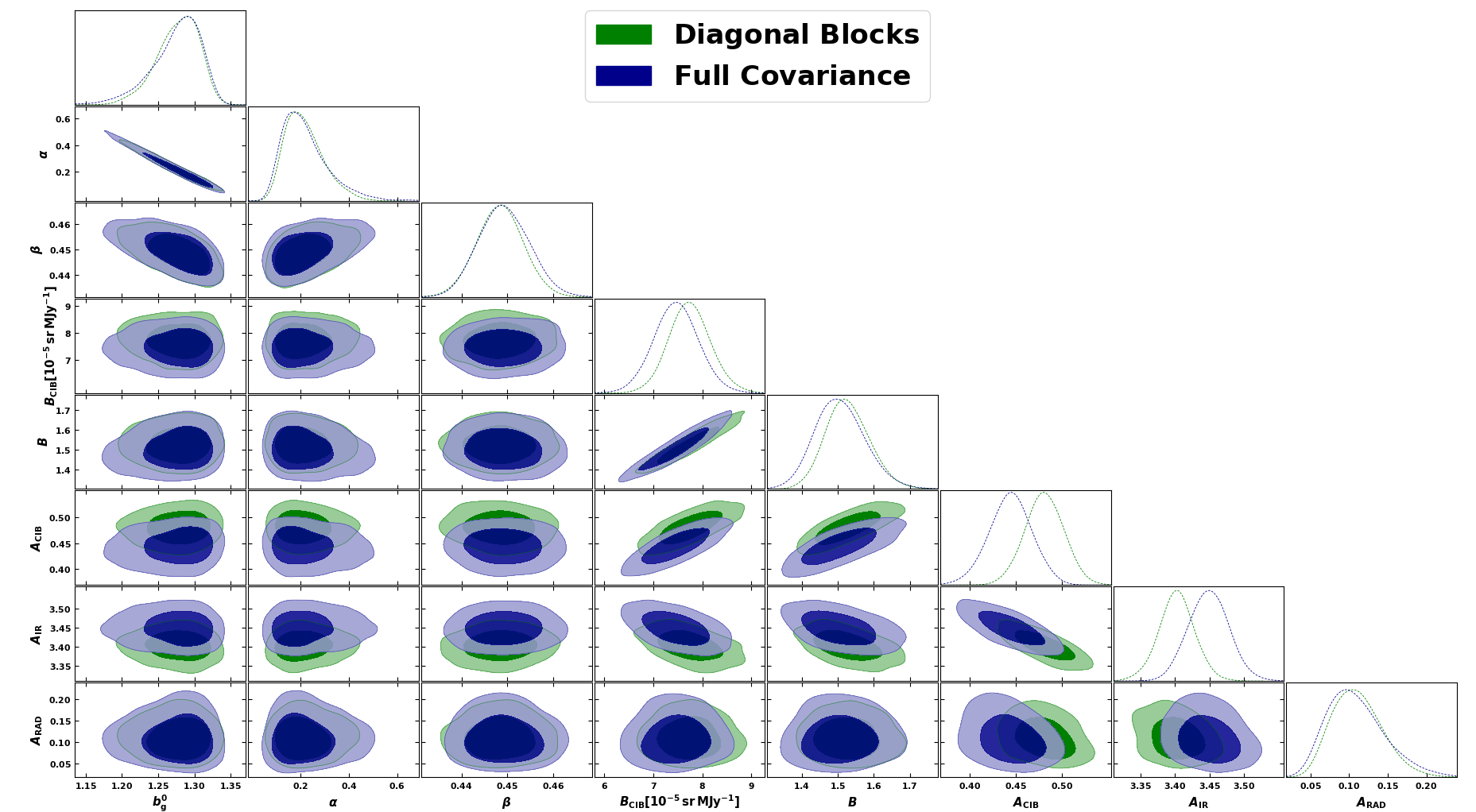}
	\caption{Posterior distributions for the eight free parameters in our model. 
	        The figure shows the joint posterior contours for all parameter 
		pairs, and the marginalised one-dimensional posterior distribution 
		for each parameter along the table diagonal. Results are shown for 
		both the full covariance and the diagonal blocks cases (Sec.~\ref{ssec:likelihood}).}
	\label{fig:posteriors}	
\end{figure*}

\subsection{modeling observational effects}
\label{ssec:pseudocl}

The power spectra predictions computed with the algorithm described above cannot be directly compared with the measurements 
presented in Sec.~\ref{sec:analysis}, because they do not include the effect of the beam convolution or the artificial mode 
coupling induced by the mask. The recipe described in~\citet{Hivon02} allows to account for these effects and to compute the 
associated pseudo-power spectrum $C^{\rm P}_{\ell}$. The pseudo power spectrum is obtained from the theoretical power spectrum 
$C_{\ell}$ using the following transformation: 
\begin{eqnarray}
\label{eq:PCL}
C^{\rm P}_{\ell} = \sum_{\ell^{\prime}} M_{\ell \ell^{\prime}} \, B^2_{\ell'} \, C_{\ell^{\prime}}. 
\end{eqnarray}
In the equation above, $B_{\ell}=\exp\left(-\ell^{2}\sigma^{2}_{\rm b}/2 \right)$ is the beam window function, where 
$\sigma_{\rm b}$ is related to the beam full width at half maximum 
$\theta_{\rm FWHM}$ by $\sigma_{\rm b}=\theta_{\rm FWHM}/\sqrt{8\ln 2}=0.00742\left(\theta_{\rm FWHM}/1^{\circ} \right)$. For 
the \textit{Planck} Compton maps we have $\theta_{\rm FWHM}=10'$, and our projected galaxy density map was degraded to 
the same resolution. The factor $M_{\ell \ell'}$ in Eq.~(\ref{eq:PCL}) is the mode-coupling matrix which is calculated as:
\begin{align}
M_{\ell_1 \ell_2}  &=  (2 \ell_2 + 1) \nonumber \\
&\times \sum_{\ell_3 }  \frac{(2 \ell_3 + 1)}{4 \pi} \, \tilde{W} _{\ell_3} \,
	{\left ( \begin{array}{ccc}
        \ell_1 & \ell_2 & \ell_3  \\
        0  & 0 & 0
       \end{array} \right )^2}, \label{PCL8}
\end{align}
where the term in round brackets is the Wigner-3j symbol, and $W_{\ell}$ is the power spectrum of the mask. 

In our implementation we employ the {\tt MASTER} code~\citep{Hivon02} to perform this 
computation\footnote{Specifically, we employ the \texttt{FORTRAN90} routines available at Prof. E. 
Komatsu webpage (\url{https://wwwmpa.mpa-garching.mpg.de/~komatsu/crl/list-of-routines.html}) to carry 
out the computation of the $M_{\ell_1 \ell_2}$ matrix.}. Notice that although our measured spectra are 
binned over the effective multipoles $\ell_{\rm eff}$, the coupling in Eq.~(\ref{eq:PCL}) is to be evaluated 
for all individual multipoles $\ell$. Hence, we perform the spectrum binning into bandpowers only on the 
final pseudo-power spectrum $C^{\rm P}_{\ell}$, and not on the simple theoretical prediction $C_{\ell}$, 
before comparing it with our measurements.

\begin{table*}
\centering
\caption{Results of our parameter estimation analysis. We quote the best-fitting estimate with associated uncertainties 
	and the corresponding 68\% confidence range. For each parameter we adopt a flat (uninformative) prior in its physical unit. Fitting results are quoted for both the diagonal blocks of covariance and the full covariance cases. The unit of $B_{\rm CIB}$ is $[10^{-5}\,{\rm sr}\,{\rm MJy}^{-1}]$ for the best-fitting values, while its prior range is $[0,2]\times {\rm sr}\,{\rm MJy}^{-1}$.}
\label{tab:estimates}		
\begin{tabular}{cccc} 		
\hline  
\hline 
Parameter     & Prior-Range &  Best-fits (Diagonal) &  Best-fits (full Cov.)   \\
\hline
$b_{\rm g}^0$ & $[0,3]$ & $1.28{\pm 0.03}$ & $1.28^{+0.03}_{-0.04}\,(\textrm{stat}) \pm{0.11}\,(\textrm{sys})$ \\

${\alpha}$ & $[0,2]$ & $0.21^{+0.09}_{-0.07}$ & $0.20^{+0.11}_{-0.07}\,(\textrm{stat}) \pm{0.10
}\,(\textrm{sys})$ \\

${\beta}$ & $[0,2]$ & $0.45\pm{0.01}$ & $0.45{\pm 0.01}\,(\textrm{stat}) \pm{0.02}\,(\textrm{sys})$  \\

$B$ & $[0,3]$ & $1.53{\pm 0.06}$ & $1.50{\pm0.07}\,(\textrm{stat}) \pm{0.34}\,(\textrm{sys})$ \\

$B_{\rm CIB}$ & $[0,2]$ & ${7.73}^{+0.43}_{-0.42}$ & ${7.46}^{+0.45}_{-0.44}\,(\textrm{stat}) \pm{ 0.58}\,(\textrm{sys})$ \\

$A_{\rm CIB}$ &$[0,10]$& $0.48 {\pm 0.02}$  & $0.44{\pm 0.02}\,(\textrm{stat}) \pm{ 0.01}\,(\textrm{sys})$ \\

$A_{\rm IR}$ &$[0,10]$ & $3.40\pm{0.03}$  & $3.45\pm{0.03}\,(\textrm{stat}) \pm{ 0.04}\,(\textrm{sys})$ \\

$A_{\rm Rad}$ & $[0,10]$ &  $0.11\pm{0.03}$ &  $0.10^{+0.04}_{-0.03}\,(\textrm{stat}) \pm{ 0.04}\,(\textrm{sys})$ \\
\hline
\end{tabular}
\end{table*}


\section{Parameter Estimation}
\label{sec:parest}

The theoretical modeling described in Sec.~\ref{sec:modelling} allows us to compute the auto- 
and cross-correlation power spectra between the Compton parameter and galaxy overdensity maps, 
provided a set of parameters is defined. The parameters include the SZ mass bias parameter $B$, 
the galaxy bias parameter defining the kernel of the projected density field (more specifically, 
its normalisation $b_{\rm g}^0$ and its redshift and multipole scaling power indices $\alpha$ and 
$\beta$), and the nuisance parameters quantifying the amplitude of the foreground contaminations. 
In this section, we describe the methodology we employ to provide constraints on these parameters 
against the power spectra measured in Sec.~\ref{sec:analysis}, and discuss the resultant estimates. 


\subsection{Methodology}
\label{ssec:likelihood}

For the rest of this work we shall fix the cosmology and only let the foreground parameters and 
the mass and galaxy bias parameters free to vary. Although the bias parameters are the main focus 
of our work, we are also interested in assessing to what extent the degeneracy with the foreground 
parameters can affect their estimation. The parameter space we explore is then eight-dimensional, 
each point of which can be expressed as a vector 
$\Theta=\{B, A_{\rm CIB}, A_{\rm IR}, A_{\rm Rad}, B_{\rm CIB}, b_{\rm g}^0, \alpha, \beta\}$.
The best-fitting 8-tuple can be determined by maximising a suitable likelihood function $\mathcal{L}$, 
or by minimising a corresponding $\chi^2$ function defined as $\chi^2=-2\,\ln{\mathcal{L}}$. As 
anticipated in Sec.~\ref{ssec:covmatr}, we want to fit our model against all of the observed auto- 
and cross-correlations at the same time. The theoretical model can then predict a theoretical vector 
$\mathbf{C}^{\rm theo}_{\ell}(\Theta)$ as defined in Eq.~(\ref{eq:vector}), which would depend this 
time on the parameter set $\Theta$. If we label by $\mathbf{C}_{\ell}^{\rm obs}$ the vector whose 
components are the spectra measured in Sec.~\ref{sec:analysis}, the likelihood function of $\Theta$ 
can be computed by using the full covariance matrix ${\rm Cov}_{\rm tot}$, defined in 
Eq.~(\ref{eq:covfull}), as:
\begin{eqnarray}
	\chi^{2}(\Theta) &=& \left[\mathbf{C}_{\ell}^{\rm obs}-\mathbf{C}_{\ell}^{\rm theo}(\Theta)\right]\,{\rm Cov}_{\rm tot}^{-1}\, \nonumber\\
	&\times & \left[\mathbf{C}_{\ell}^{\rm obs}-\mathbf{C}_{\ell}^{\rm theo}(\Theta)\right]^{\rm T},
\end{eqnarray}
where ${\rm Cov}_{\rm tot}^{-1}$ denotes the inverse of the covariance matrix, and the theoretical 
vector is made of the pseudo-power spectra computed with Eq.~(\ref{eq:PCL}). The procedure we follow 
for inverting the full covariance matrix is detailed in Appendix~\ref{sec:invertcov}.

We also consider a parameter estimation performed without using the full covariance matrix, 
reverting instead to the computation of three independent $\chi^2$, one for each spectra, and 
summing their contribution together:
\begin{equation}
	\chi^2 = \chi_{\rm gg}^2 + \chi_{\text{g}y}^2 + \chi_{yy}^2.
\end{equation}
By construction, this is equivalent to using the full covariance matrix but setting the non-diagonal 
building blocks to zero (i.e. neglecting the correlation between different spectra). We shall refer 
to this method as the ``diagonal blocks'' case, in contrast to the ``full covariance'' case, which also 
takes into account the non-diagonal blocks in Eq.~(\ref{eq:covfull}).

We explore the parameter space using a Markov Chain Monte Carlo (MCMC) approach. We employ the 
\texttt{Python} \textsc{emcee} package\footnote{\url{https://emcee.readthedocs.io/en/stable/}.}~\citep{Foreman}, 
which allows to set priors on the parameters and specify the number of chains. We employed 100 chains 
with a total number of effective steps of 50000 after burn-in removal and chain thinning. The thinning factor was 
chosen as half the auto-correlation time, which represents the number of steps taken by each chain before reaching a position that is 
uncorrelated from the starting one, so that our thinned chains can be considered independent draws of parameter values from their 
posterior distributions; the large number of points per chain ($>50$) still available after thinning ensures that our 
chains have reached convergence. We then use the \texttt{Python} \textsc{GetDist} 
package~\citep{lewis19}\footnote{\url{https://getdist.readthedocs.io/en/latest/}.}
to retrieve the final posterior distributions on the parameters, which are plotted in 
Fig.~\ref{fig:posteriors} for both the full covariance and the diagonal blocks cases.
The final estimates on the fitted parameters for both cases, together with their uncertainties and 
initial priors, are summarised in Table~\ref{tab:estimates}. The associated best-fit predictions 
for the gg, g$y$ and $yy$ power spectra are overplotted to the measured data points in Fig.~\ref{fig:spectra}. 

We stress that the parameter errors extracted from their posterior distributions only quantify 
their statistical uncertainty. For the case of the full covariance, in Table~\ref{tab:estimates} we quote, 
in addition, our estimates for the systematic uncertainties affecting the parameters. These systematic 
errors were obtained by modifying the WISE redshift distribution shown in Fig.~\ref{fig:dndz-fit}. 
A more detailed description is provided in Appendix.~\ref{sec:pz_systematics}, together with a general discussion 
on the possible sources of systematics affecting our analysis.


\begin{table*}
\centering	
\caption{{Comparison between different constraints for the tSZ mass bias, expressed as $1-b_{\rm H}$, including the one obtained in this work. For each case we report the main observable(s) employed in the analysis, the corresponding specific survey/instrument and the considered mass range, although we redirect to the corresponding references for details. Constraints above the horizontal line all involve the use of the tSZ Compton maps. ``WL'' refers to weak lensing for brevity.}}
\label{tab:Bias_comparison}
\renewcommand{\arraystretch}{1.1}
\begin{tabular}{lclll}
\hline 
	Observables & Survey & $1-b_{\rm H}$ &Mass Range $[h^{-1}{\rm M}_{\odot}]$& Reference\\
\hline
	tSZ + galaxy density field & \textit{Planck}, WISE & 0.67  $\pm$  0.03 &$5\times 10^{11} - 5\times 10^{15}$ & This work\\
	tSZ + cluster catalogues    & \textit{Planck}	   & 0.60 $\pm$ 0.05 & $10^{11} - 5\times 10^{15}$& \cite{rotti21} \\
				    & 			   & 0.85 $\pm$ 0.04 & $10^{11} - 5\times 10^{15}$& \cite{rotti21}\footnote{This result was obtained by removing resolved clusters.} \\
	tSZ tomography & \textit{Planck}, SDSS & 0.79 $\pm$ 0.03 &$10^{11} - 5\times 10^{15}$  &\cite{chiang20}\\				    
	tSZ + WL & \textit{Planck}, HSC & $0.73^{+0.08}_{-0.13}$ & $10^{10} - 10^{16}$&\cite{Makiya2019} \\	
	tSZ + X-ray + WL  & \textit{Planck}, ROSAT & 0.71 $\pm$ 0.07 & $10^{13} - 10^{16}$ &\cite{Hurier2018} \\	
	tSZ + WL & \textit{Planck}, HSC &  $0.63^{+0.04}_{-0.09}$ &$2\times 10^{13} - 10^{16}$& \cite{Ken-Osato2019} \\
	tSZ + CMB & \textit{Planck} & 0.62 $\pm$ 0.05 &$10^{14} - 10^{15}$& \cite{Salvati-bias} \\
	tSZ + CMB & \textit{Planck} & 0.58 $\pm$ 0.06&$10^{11} - 5\times 10^{15}$& \cite{Bolliet18}\\
	tSZ + WL & \textit{Planck}, SDSS & 0.74 $\pm$ 0.07 & $3.5\times 10^{13} - 2\times 10^{14}$& \cite{Hurier2017} \\
	tSZ + galaxy density field & \textit{Planck}, 2MASS & 0.65 $\pm$ 0.04 &$10^{10} - 10^{16}$ & \cite{Makiya2018} \\	
\hline
	WL + cluster counts & \textit{Planck} & 0.62 $\pm$ 0.03 & $2\times 10^{14} - 10^{15}$& \cite{planck18} \\ 	
	WL & ACT & $0.74^{+0.13}_{-0.12}$ &$1.3\times 10^{14} - 6\times 10^{14}$& \cite{Miyatake2018} \\
	WL & \textit{Planck} & 0.71 $\pm$ 0.10 &$ 10^{13} - 10^{16}$ & \cite{Zubeldia2019}\\	
	WL & \textit{Planck} & 1.01 $\pm$ 0.19 & $5\times 10^{14} - 7\times 10^{15}$& \cite{Planck2015-XXIV}\\	
	WL & CCCP & 0.76 $\pm$ 0.08 &$5\times 10^{14} - 2.5\times 10^{15}$ & \cite{Hoekstra2015}\\ 
	WL & Weighing the Giants & 0.69 $\pm$ 0.07 & $2\times 10^{14} - 3\times 10^{15}$& \cite{Linden-bias}\\
\hline
\end{tabular}
\end{table*}

\subsection{Discussion}
\label{ssec:discussion}
We discuss in this section the results obtained from the parameter estimation analysis, beginning with the 
difference between the diagonal blocks and the full covariance cases. The contours in Fig.~\ref{fig:posteriors} 
show that the results are generally compatible: the only posterior distributions that show a mild tension between 
the two cases are those involving the clustered cosmic infrared background $A_{\rm CIB}$ and the amplitude of 
the infrared source contamination $A_{\rm IR}$. However, these deviations are always within one sigma. We 
conclude therefore that the usage of the full covariance matrix, which takes into account the cross-correlation 
between different observed spectra, produces a consistent result with respect to the case of considering only 
the corresponding covariances. This could be expected as the off-diagonal blocks of the full covariance matrix 
provide only a second-order contribution with respect to the diagonal ones.

We observe hints of anti-correlation between the parameters $A_{\rm CIB}$ and $A_{\rm Rad}$, which can be 
expected as the associated foreground spectra have a very similar multipole dependence. The tight 
anti-correlation between the galaxy bias normalization $b^0_{\rm g}$ and the slope of the redshift dependence 
$\alpha$ is simply a result of our chosen functional form for $b_{\rm g}$ (Eq.~(\ref{eq:bz})); similar 
considerations apply to the joint posterior distribution between $b^0_{\rm g}$ and $\beta$. A positive 
correlation is found instead between $A_{\rm CIB}$ and $B_{\rm CIB}$, which is understandable as they both 
gauge the level of CIB contamination, although in different power spectra. We also observe a positive correlation 
between $B$ and $B_{\rm CIB}$: a higher value of $B_{\rm CIB}$ requires a lower contribution from the tSZ power 
spectrum to fit our data points, thus favouring a lower Compton parameter amplitude which can be achieved via a 
higher bias $B$. 

We consider now the best-fit values quoted in Table~\ref{tab:estimates}. The nuisance parameters controlling 
the foreground amplitudes in the $yy$ auto-spectrum have also been constrained in previous 
works~\citep{Makiya2018, Bolliet18, rotti21}. Our $A_{\rm CIB}$ and $A_{\rm Rad}$ estimates 
are consistent with the findings of~\citet{Makiya2018} and~\citet{rotti21} respectively, while we find a 
higher value for the $A_{\rm IR}$ parameter instead. However, the actual, individual values of the nuisance 
parameters are not of particular interest, as different amplitudes in Eq.~(\ref{eq:clforegrounds}) can lead 
to the same observed power spectrum. What matters in this context is the overall foreground contamination 
from all these sources combined. From the third panel of Fig.~\ref{fig:spectra} we see that the foreground 
contribution dominates the $yy$ auto-correlation at multipoles $\ell\gtrsim 200$. This clearly shows the 
necessity of including the nuisance parameters in our fit and that their fitted values allow recovering 
the observed spectral amplitude at the smallest scales. 

In addition, we first provide an estimate of $B_{\rm CIB}$, which controls the CIB contamination to the 
cross-correlation of the galaxy field with the Compton parameter map. The best-fit value is of order 
$10^{-5}\,\text{sr}\,\text{MJy}^{-1}$ that occupies the lower end of the prior range ($[0,2]\times {\rm sr}\,{\rm MJy}^{-1}$), and the associated contamination to the g$y$ cross-correlation 
allows us to recover the amplitude of the measured power spectra at large scales, as clearly shown in the second 
panel of Fig.~\ref{fig:spectra}. Similarly, the $B_{\rm CIB}$ contribution is particularly relevant also at 
very small scales, where it has an amplitude larger than the two-halo term and enables our theoretical prediction 
to match the observed spectral amplitude. This proves the importance of accounting for the CIB contribution 
when tSZ is cross-correlated with galaxy catalogues at $z\sim 1$ or above.

We consider now the constraints obtained on the tSZ mass bias parameter, which is the main goal of our work. 
As already mentioned, several previous works have fitted for $b_{\rm H}$ against different 
observables~\citep[see, e.g. Table 1 in][]{Ma2015}, as summarised in Table~\ref{tab:Bias_comparison}. With our 
analysis we obtain the estimate $B =1.50{\pm 0.07}\,(\textrm{stat}) \pm{0.34}\,(\textrm{sys})$, 
which corresponds to $1-b_{\rm H}=0.67\pm 0.03\,({\rm stat})\pm 0.16\,({\rm sys})$; the latter value 
is also reported in Table~\ref{tab:Bias_comparison}. Our finding is largely in agreement with other estimates 
and is particularly consistent with~\citet{Makiya2018}, who also considered the cross-correlation between tSZ  
and galaxy density field. Although results from hydrodynamical simulations suggest that the tSZ mass underestimates 
the total cluster mass by only 5\% to 20\%~\citep{Meneghetti2010,Truong2018,Angelinelli,Ansarifard,Pearce,Barnes,Gianfagna}, to solve the tension 
between cluster-based and CMB-based estimates on cosmological parameters, higher bias values are 
required~\citep{Planck2015-XXIV}. The results from~\citet{rotti21} show that by 
excising the contribution of detected/resolved clusters from the \textit{Planck} $y$ map, a power spectrum 
analysis yields $1-b=0.85\pm0.04$, in agreement with simulations, while the inclusion of massive clusters leads 
to the estimate $1-b=0.60\pm0.05$, which is lower than our findings. The reference points out that this can 
result from a possible mass dependence of the bias or CIB contaminations, with novel data required to 
provide a deeper understanding. Our constraint suggests that the tSZ mass underestimates the true cluster mass by 
$\sim 33\%$, thus corroborating the hypothesis that the mass bias is higher than the results favoured by 
numerical simulations alone. 

Finally, the linear galaxy bias amplitude at $z=0$ and $\ell=117$ is constrained to 
$b_{\rm g}^0=1.28^{+0.03}_{-0.04}\,(\textrm{stat}) \pm{0.11}\,(\textrm{sys})$,
while the power index for its redshift and multipole dependences are 
$\alpha=0.20^{+0.11}_{-0.07}\,(\textrm{stat}) \pm{0.10}\,(\textrm{sys})$ and $\beta=0.45{\pm 0.01}\,(\textrm{stat}) \pm{0.02}\,(\textrm{sys})$. From the first panel in Fig.~\ref{fig:spectra} 
we see that the inclusion of a redshift and multipole dependence in the galaxy bias allows to recover the measured gg 
correlation at small scales, but tends to underestimate the spectral amplituce at the largest scales. This could be a result 
of our theoretical modeling, and could possibly be solved by opting for a full halo occupation distribution (HOD) approach 
instead of the halo model. For our parametrisation, the value of $\alpha$ denotes a mild redshift dependence, which results in a 
mean value for the galaxy bias across WISE redshift range (at the reference multipole) of $\bar{b}_{\rm g}\simeq 1.37$. 
It is indeed expected to obtain a higher linear galaxy bias with increasing redshift for an approximately magnitude-limited 
galaxy sample, in agreement with several observational and theoretical 
studies~\citep{Somerville01_linearbias,Gazta_linearbias,Crocce_linearbias,Merson2019_linearbias}.
We can compare our findings with previous estimates of the galaxy bias from works employing WISE data. In~\citet{Ferraro2015} a similar 
functional form for the redshift evolution of the galaxy bias was considered, with a fixed exponent $\alpha=1$ and a fitted normalisation 
$b_{\rm g}^0=0.98 \pm 0.10$. The same reference also considered a model with a constant bias, which yielded the estimate $\bar{b}_{\rm g}=1.41\pm0.15$, 
which is compatible with our redshift-averaged value of $1.37$. The works in~\citet{hill16} and~\citet{ferraro16} favour instead the mean 
value $b_{\rm g}=1.11 \pm 0.08$, which is lower than our normalisation at $z=0$. Finally, we point out that our overall error estimate for 
$b_{\rm g}^0$, considering both the statistical and the systematic contributions, is comparable with the uncertainties quoted by those studies.


\subsection{Effective mass range}
\label{ssec:massrange}
In order to better understand the best-fit values presented in Sec.~\ref{ssec:discussion}, it is interesting 
to investigate what mass range has the highest statistical weight in constraining our model. To this aim, we 
follow the formalism presented in~\citet{rotti21}, which allows to estimate the mean mass that contributes to 
the computation of the power spectrum for each multipole $\ell$. The computation is performed separately for the 
one-halo and the two-halo term via a weighted average over mass and redshift. For the generic $AB$ 
cross-correlation, the mean mass contributing to the one-halo term reads:
\begin{equation}
	\label{eq:m1h}	
	\langle M \rangle_{\ell}^{AB, {\rm 1h}} = \dfrac{\int\text{d}z \int\text{d}M \, M\, f_{\ell}^{AB}(M,z)}{ \int\text{d}z \int\text{d}M \,f_{\ell}^{AB}(M,z)},
\end{equation}
where it is understood the integrals are to be evaluated within the chosen ranges $[z_{\rm min},z_{\rm max}]$ 
and $[M_{\rm min},M_{\rm max}]$, and we have introduced the short-hand notation:
\begin{equation}
\label{eq:fab}
f_{\ell}^{AB}(M,z) = \dfrac{c\chi^2(z)}{H(z)}\,\dfrac{\text{d}n}{\text{d}M}(M,z)\,A_{\ell}(M,z)\,B_{\ell}(M,z).
\end{equation}
The weighing factor $f_{\ell}^{AB}(M,z)$ is obtained as the product of the comoving volume element, the halo 
mass function and the Fourier transforms $A_{\ell}(M,z)$ and $B_{\ell}(M,z)$. Similarly, for the two-halo 
term we have:
\begin{equation}
	\label{eq:m2h}	
	\langle M^2 \rangle_{\ell}^{AB, {\rm 2h}} = \dfrac{\int\,\text{d}z\, q_{\ell}(z)\, G(M A_{\ell}) \, G(M B_{\ell}) } { \int\,\text{d}z\, q_{\ell}(z)\, G(A_{\ell})\, G(B_{\ell})},
\end{equation}
where
\begin{equation}
	q_{\ell}(z) = \dfrac{c\chi^2(x)}{H(z)}\,P^{\rm{lin}}_{\rm{m}}\left(\frac{\ell +1/2}{\chi(z)},z\right),
\end{equation}
and
\begin{equation}
	G(x) =  \int\,\text{d}M \,x\, \dfrac{\text{d}n}{\text{d}M}(M,z)\,b(M,z),
\end{equation}
with $P^{\rm{lin}}(k,z)$ the linear matter power spectrum and $b(M,z)$ the halo bias (same as in 
Eq.~\ref{eq:2halo}). 

\begin{figure}
\centering
\includegraphics[width=8cm]{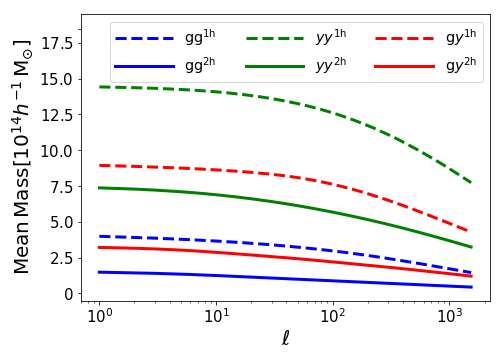}
\caption{The mean halo mass, as a function of $\ell$, which mostly contributes to the 
	computation of the auto- and cross-power spectra, calculated using Eq.~(\ref{eq:m1h}) for the 
	one-halo term and Eq.~(\ref{eq:m2h}) for two-halo term. Results are shown for all our correlations 
	cases, marked as gg, g$y$ and $yy$.}
\label{fig:mrange}
\end{figure}

The resulting mean masses $\langle M \rangle_{\ell}^{AB, {\rm 1h}}$ and $\sqrt{\langle M^2 \rangle_{\ell}^{AB, {\rm 2h}}}$
are plotted as a function of the multipoles in Fig.~\ref{fig:mrange}, for the gg, g$y$ and $yy$ cases.
We see that in all cases the mean mass decreases with $\ell$, thus suggesting the contribution from 
lower mass clusters dominates the computation of the power spectra at smaller scales, as expected. 
In general, the multipole dependence is stronger for the one-halo term than for the two-halo term, with the 
former being dominated by higher masses. When comparing different correlation cases, we notice the 
$yy$ correlation is dominated by higher masses compared to the gg correlation, with the g$y$ case 
in between. On average, the mean mass that mostly contributes to our $yy$ measurement is of order 
$10^{15}\,h^{-1}M_{\odot}$; this corresponds to the scale of massive clusters, which provide indeed 
the strongest signal in the $y$ map. The gg power spectrum, instead, mainly receives its contribution 
from masses below $4\times 10^{14}h^{-1}{\rm M}_{\odot}$. This may be linked to a difference in the 
nature of the projected galaxy density field and the Compton parameter as LSS tracers. While galaxy 
overdensities can be more readily linked to the underlying dark matter distribution, the hot gas 
responsible for the tSZ effect requires a higher gravitational potential; the latter is only reached at the 
peaks of the dark matter distribution, where the inter-galactic medium can be ionised and local galaxies 
virialise into galaxy clusters. We can expect therefore higher halo masses to provide the main contribution 
to the $yy$ spectrum.  

In Table~\ref{tab:Bias_comparison}, we list the mass range explored by previous works in the literature. We notice that 
the effective mass range derived in this section is consistent with the ones employed 
by~\citet{Hoekstra2015} and~\citet{Linden-bias}; our estimate for the cluster mass bias is compatible with 
the results provided by those works. 

\subsection{Possible systematics}

This section is dedicated to a review of the most likely sources of systematics that can affect our findings, 
related to both our data set and the methodology adopted in this paper.

Regarding our data set, the strongest source of systematic is the choice of a sensible redshift distribution
to describe our selected WISE galaxy sample. As already commented in Sec.~\ref{ssec:wise}, we adopt the redshift 
distribution derived in~\citet{Yan2013} for the sample of sources detected in the W1 band with $S/R>7$. This 
distribution may not be entirely representative of the galaxy sample employed in this analysis, due to the 
cut we applied on WISE data. Besides, the redshift distribution was derived by cross-matching WISE detections 
with optical SDSS sources; this procedure may not be adequate for WISE higher redshift detection, which could 
be missed by the SDSS selection. In order to take these issues into account, we follow the strategy adopted 
in~\citet{Ferraro2015}: we repeat the full analysis by considering two modified versions of the WISE redshift 
distribution, obtained by shifting the fitted $p_{\rm s}(z)$ function by $\Delta z=\pm0.1$. These 
distributions are compared in Fig.~\ref{fig:offset_dndz}, while the results of the corresponding parameter estimation are 
shown in Fig.~\ref{fig:offset_posteriors} (for simplicity we moved these results to Appendix~\ref{sec:pz_systematics}); 
the best-fit parameter values are quoted in Table~\ref{tab:offset_estimates}. For each parameter, we take the largest 
of the two offsets between these new best-fits, and the ones obtained from our fiducial choice for the $p_{\rm s}(z)$ 
distribution, as the systematic uncertainty on that parameter. The resultant systematic error bars are quoted together 
with the statistical errors in Table~\ref{tab:estimates}.

Another possible source of systematics is our use of a halo model to predict theoretically the measured correlations. 
In this context, using a full HOD model could provide a better fit to the gg power spectra, especially at large 
scales. This approach was employed for example in~\citet{Makiya2018}. We feel, however, that the use of an HOD model 
may provide a better fit at the expense of increasing the number of free parameters; in the current analysis we 
prefer to keep using a halo model with less but more physically representative parameters (the halo mass bias and the 
galaxy bias). In particular, it is interesting to provide new constraints on the galaxy bias parameter using the 
cross-correlations obtained from our WISE projected density maps; the galaxy bias would not appear in our modeling 
if we employed an HOD approach. 

Finally, for our modeling of the cluster pressure profile we employ the UPP form from Eq.~(\ref{eq:upp}), using the 
best-fit parameters obtained in~\citet{Planck2013}; the latter were estimated on a set of 62 clusters with masses $M_{500}>2\times10^{14}h^{-1}{\rm M}_{\odot}$ at $z<0.45$. As our analysis extends to higher redshifts and includes lower masses, 
it is legit to argue that our chosen UPP parametrization may not be suitable for objects with lower masses and higher 
redshifts. Adaptations of the UPP form to accommodate mass and redshift dependence have been explored for example 
in~\citet{battaglia12} and~\citet{lebrun15}. Nonetheless, the analysis we described in Sec.~\ref{ssec:massrange} 
proves that the main masses contributing to our $yy$ spectrum are of order $10^{15}\,h^{-1}{\rm M}_{\odot}$; for these high 
masses, the UPP fitted over the resolved \textit{Planck}-detected clusters is adequate. The introduction of a mass 
and redshift dependence in the UPP parameters would introduce a further complication in our modeling, and although 
it could be tackled by future studies, it goes beyond the scope of the current analysis.

To summarise this discussion, we do acknowledge that our results are affected by systematic issues. The most relevant one, 
at the data level, is the choice of the redshift distribution describing our WISE sample; as commented above, the choice
of offset versions for the $p_{\rm s}(z)$ allows to provide an estimate for the additional systematic component in our 
error bars. We choose a substantial offset of $\Delta z=\pm0.1$, and take the offset between the extreme cases as a 
measurement of our systematic errors; it is then reasonable to believe that this additional uncertainty is likely 
overestimated (at least as far as the choice of a redshift distribution is concerned). Hence, even though we do not 
explicitly quantify the systematic errors stemming from the other two points described in this section, 
our conservative apporach enables us to consider the quoted error bars as representative of the overall systematic uncertainty 
affecting our analysis.


\section{Conclusions} 
\label{sec:conclusions}

The current work aimed at providing novel constraints on the cluster mass bias parameter $B$, 
which quantifies the deviation of the tSZ-estimated cluster mass from the actual cluster mass. The difference between the two masses is due to the assumption of hydrostatic equilibrium in the ICM and other systematics affecting the determination of the underlying mass proxies. 
Although this task has already been tackled by previous work, no definitive conclusion has been reached about the value of $B$. The uncertainty on $B$ is one of the major 
issues hindering the effective use of cluster number counts as a cosmological probe. 

In this work, we fitted for the bias parameter by studying the correlation between the Sunyaev-Zel'dovich effect and the galaxy overdensity field. The former is quantified by an all-sky map of the Compton parameter published by the Planck Collaboration. 
The latter is obtained by projecting on the sky a galaxy catalogue acquired with the WISE infrared satellite. A proper mask was overlaid to these maps to excise regions affected by Galactic foregrounds or noisy pointings. With this data set, we measured the power spectra quantifying the Compton parameter auto-correlation ($yy$), the 
galaxy density auto-correlation (gg) and the cross-correlation between the two (g$y$). 
We made use of the \texttt{PolSpice} package, which allows computing power spectra of 
sky maps with customised masks, and also output the cross-correlation between pairs of 
multipoles. To maximise the statistical information encoded in our data set, we 
joint the three spectra in a unique vector and computed its full covariance, which also 
includes the correlations between different spectra in its non-diagonal terms. 

Our theoretical prediction for the observed spectra was based on a halo model. We detailed how we computed the Fourier transforms of the Compton parameter and the galaxy field and used them to evaluate the one-halo and two-halo terms. The hydrostatic mass bias parameter is a 
key ingredient in the modeling of the Compton parameter Fourier modes. Similarly, the 
Fourier transform of the galaxy density field is dependent on the linear galaxy bias, 
which controls the amplitude of the redshift distribution of the observed sources.
We allowed such bias to depend on both redshift and scale and parametrised it in terms of its amplitude 
$b_{\rm g}^0$ at $z=0$ and on the respective power-law indices $\alpha$ and $\beta$.
In our modeling, we also included the effect of foreground residuals in the Compton map, 
which affect its auto-correlation and its cross-correlation with the galaxy field. 
Eventually, we obtained a recipe to predict each correlation starting from a set of model 
parameters, including the hydrostatic mass bias $B$, the galaxy bias parameters $b_{\rm g}^0$, $\alpha$ and $\beta$,
and a set of nuisance parameters that quantify the foreground contamination in our measurements. 

We derived the posterior probability distributions for these parameters using an MCMC 
approach implemented with the \texttt{Python} \textsc{emcee} package. As a sanity check, we also repeated 
the fit by neglecting the non-diagonal terms in the full covariance (i.e. by joining \textit{a posteriori} 
the gg, g$y$ and $yy$ likelihoods), which yielded estimates close to the full-covariance case.
While the posterior probability distributions quantified the statistical errors on the parameter estimates, we 
also evaluated additional systematic uncertainties; the latter were computed as the maximum offsets between 
these best-fit values and the ones obtained by adopting different redshift distributions to model our 
galaxy sample. Specifically, we considered two additional versions of the fiducial $p_{\rm s}(z)$ distribution
obtained by shifting the baseline redshift by $\Delta z=\pm0.1$. We showed this is an important variation 
in the $p_{\rm s}(z)$, ensuring the resulting uncertainty is quite conservative and sufficient to include
other possible sources of systematics in our analysis (e.g. the choice of an halo model, or the use of a 
universal pressure profile). 

The joint distributions between parameter 
pairs did not show any strong degeneracy between the nuisance parameters and the parameters of most interest 
($B$, $b_{\rm g}^0$, $\alpha$, $\beta$) except for the case of $B$ and $B_{\rm CIB}$. We observed degeneracy though 
between $b_{\rm g}^0$ and $\alpha$, which is expected as $\alpha$ controls the slope of the redshift dependence. 
The final best-fit estimates are $B =1.50{\pm 0.07}\,(\textrm{stat}) \pm{0.34}\,(\textrm{sys})$, corresponding to 
$1-b_{\rm H}=0.67\pm 0.03\,({\rm stat})\pm 0.16\,({\rm sys})$, 
$b_{\rm g}^0=1.28^{+0.03}_{-0.04}\,(\textrm{stat}) \pm{0.11}\,(\textrm{sys})$, and the power index for its redshift 
and multipole dependences are $\alpha=0.20^{+0.11}_{-0.07}\,(\textrm{stat}) \pm{0.10}\,(\textrm{sys})$ and 
$\beta=0.45{\pm 0.01}\,(\textrm{stat}) \pm{0.02}\,(\textrm{sys})$. These results, together with the constraints 
for the foreground coefficients, prove effectiveness in reproducing the observed power spectra. 

We find a linear galaxy bias normalisation in broad agreement with the estimates found in previous works, with an increase 
in the precision as far as the statistical error bar is concerned. The small statistical uncertainty we obtain for $b_{\rm g}^0$ 
is a result of our careful treatment of systematic contaminations from CIB and other foregrounds when modeling the reconstructed power spectra.
We find a moderate bias dependence on the multipole, with smaller scales favouring a larger bias and a mild redshift dependence.   
 
Finally, our estimate for the mass bias $B$ suggests a $\sim33\%$ decrement of the tSZ mass with respect to the true cluster mass. 
This value is larger than estimates from numerical simulations, but it agrees with previous analyses that 
exploited this type of cross-correlation studies (Table~\ref{tab:Bias_comparison}).
This large value for the bias helps releasing the tension between CMB-based and 
cluster-based constraints of cosmological 
parameters~\citep[e.g. $\sigma_8\Omega_m^{0.3}$,][]{planck13_XX}. 

This type of cross-correlation analysis can be applied to future data sets, which 
will allow improving our understanding of the halo warm-hot gas physics~\citep{pandey20}.
Improved constraints of the bias can be obtained, for instance, with the next generation of galaxy 
surveys, e.g. Vera C. Rubin Observatory (LSST; \citealt{LSST2002}), Euclid~\citep{Euclid-collaboration} and DESI~\citep{DESI-collaboration}, 
and CMB missions, such as the LiteBird~\citep{LiteBIRD-collaboration} and CMB Stage-4 
experiments~\citep{CMB-S4}.


\section*{Acknowledgements} 
We would like to thank the anonymous referee for providing a very useful report. We also thank Oluwayimika Akinsiku, Boris Bolliet, Eiichiro Komatsu, Ryu Makiya and Alexander van Engelen for helpful 
discussion. A.I. acknowledges funding by the National Research Foundation (NRF) with grant no.120385. D.T. acknowledges the 
support from the National Science Foundation of China with Grant no.~12150410315, Chinese Academy of Sciences President's International Fellowship Initiative with Grant N. 2020PM0042, and 
the South African Claude Leon Foundation that partially funded this work. Y.Z.M. acknowledges the support of NRF-120385 and
NRF-120378. WMD acknowledges the support from ``Big Data with Science and Society'' UKZN Research Flagship Program.

\section*{Software and data}

For the analysis presented in this manuscript we made use of the following software:
\software{HEALPix~\citep{Gorski2005}, PolSpice~\citep{challinor11}, MASTER~\citep{Hivon02}, 
 emcee~\citep{Foreman} and GetDist~\citep{lewis19}}.

The data underlying this article are publicly available. The \textit{Planck} Compton 
parameter maps are available in the Planck Legacy Archive at 
\url{http://pla.esac.esa.int/pla}. The WISE source catalogue can be 
queried in the NASA/IPAC Infrared Science Archive (IRSA) at 
\url{https://irsa.ipac.caltech.edu/Missions/wise.html}.

\appendix


\section{Inversion of the full covariance matrix}
\label{sec:invertcov}

We detail in the following the procedure we adopted to invert the full covariance matrix defined in Eq.~(\ref{eq:covfull}). 
In order to simplify the notation, we can re-label its six independent building blocks as:
\begin{eqnarray}
	a\equiv{\rm Cov}^{\rm gg,gg} \qquad  b &\equiv& {\rm Cov}^{{\rm gg,g}y} \qquad c\equiv{\rm Cov}^{{\rm gg},yy} \nonumber \\
	d\equiv {\rm Cov}^{{\rm g}y, {\rm g}y} \qquad e &\equiv& {\rm Cov}^{{\rm g}y, yy} \qquad f\equiv{\rm Cov}^{yy,yy}, 
\end{eqnarray}
so that the full covariance matrix reads:
\begin{equation}
{\rm Cov}_{\rm tot}=\left[
\begin{array}{ccc}
a & b & c \nonumber \\
b^{\rm T} & d & e \nonumber \\
c^{\rm T} & e^{\rm T} & f \nonumber \\
\end{array}%
\right]. \label{eq:label-2}
\end{equation}

In the trivial case of considering only the diagonal blocks and setting the non-diagonal 
ones to zero, $b=0$, $c=0$, $e=0$, the inverse covariance can be computed as:
\begin{equation}
{\rm Cov}_{\rm tot}^{-1}
 = \left[
\begin{array}{ccc}
a^{-1} & 0 & 0 \\
0 & d^{-1} & 0 \\
0 & 0 & f^{-1} \\
\end{array}%
\right],  \nonumber \\ 
\label{eq:invdiag}
\end{equation}
i.e. by simply inverting the diagonal blocks. This is the inverse covariance we employ 
for the parameter estimation when neglecting the cross-correlation between different 
power spectra (the fourth columns in Table~\ref{tab:estimates}). 

In the more general case, it is still possible to partition the full covariance matrix into four blocks as:
\begin{equation}
\label{eq:blocks}
{\rm Cov}_{\rm tot}=\left[
\begin{array}{cc}
\mathbf{A} & \mathbf{B} \\
\mathbf{B}^{\rm T} & \mathbf{D} \\
\end{array}
\right],
\end{equation}
where 
\begin{equation}
	\mathbf{A} = \left[
\begin{array}{cc}
a & b  \\
b^{\rm T} & d  \\
\end{array}%
\right], \nonumber \\
\mathbf{B} = \left[
\begin{array}{c}
c  \\
e  \\
\end{array}%
\right], \nonumber \\
\mathbf{D} = f .
\end{equation}
Matrices with the general form expressed by Eq.~(\ref{eq:blocks}), where $\mathbf{A}$, 
$\mathbf{B}$, $\mathbf{D}$ have arbitrary size, can be inverted as:
\begin{equation}
\label{eq:inverse}
{\rm Cov}_{\rm tot}^{-1}=\left[
\begin{array}{cc}
\mathbf{A} & \mathbf{B} \\
\mathbf{B}^{\rm T} & \mathbf{D} \\
\end{array}
\right]^{-1} = \left[
\begin{array}{cc}
	\mathbf{W} & \mathbf{X} \\
	\mathbf{Y} & \mathbf{Z} \\
\end{array} 
\right], 
\end{equation}
where:
\begin{eqnarray}
	\label{eq:invterms}
	\mathbf{W} &=& \mathbf{A}^{-1}+\mathbf{A}^{-1}\mathbf{B}\left(\mathbf{D}-\mathbf{B}^{\rm T}\mathbf{A}^{-1}\mathbf{B} \right)^{-1}\mathbf{B}^{\rm T} \mathbf{A}^{-1} \nonumber \\
	\mathbf{X} &=& -\mathbf{A}^{-1}\mathbf{B}\left(\mathbf{D}-\mathbf{B}^{\rm T}\mathbf{A}^{-1}\mathbf{B} \right)^{-1} \nonumber \\
	\mathbf{Y} &=& -\left(\mathbf{D}-\mathbf{B}^{\rm T}\mathbf{A}^{-1}\mathbf{B} \right)^{-1}\mathbf{B}^{\rm T}\mathbf{A}^{-1} \nonumber \\
	\mathbf{Z} &=& \left(\mathbf{D}-\mathbf{B}^{\rm T}\mathbf{A}^{-1}\mathbf{B} \right)^{-1} .
\end{eqnarray}
In the derivation above it is implied that $\mathbf{A}$, $\mathbf{D}$ and $\mathbf{D}-\mathbf{B}^{\rm T}\mathbf{A}^{-1}\mathbf{B}$ 
must be square, invertible matrices. The diagonal case can be recovered by setting $\mathbf{B}=0$. The only non-trivial term 
now is the inverse $\mathbf{A}^{-1}$, which can be computed as follows:
\begin{equation}
A^{-1}=
\left[
\begin{array}{cc}
f & g \\
h & i \\
\end{array}
\right],
\end{equation}
where:
\begin{eqnarray}
\label{eq:inv_A_terms}
f &=& a^{-1}+a^{-1}b\left(d-b^{\rm T}a^{-1}b \right)^{-1}b^{\rm T} a^{-1} \nonumber \\
g &=& -a^{-1}b\left(d-b^{\rm T}a^{-1}b \right)^{-1} \nonumber \\
h &=& -\left(d-b^{\rm T}a^{-1}b \right)^{-1}b^{\rm T}a^{-1} \nonumber \\
i &=& \left(d-b^{\rm T}a^{-1}b \right)^{-1}.
\end{eqnarray}

In our $\chi^2$ computation we implement the transformations in Eqs.~(\ref{eq:inverse}) to~(\ref{eq:inv_A_terms})
in order to minimize possible numerical errors deriving from the inversion of a high rank matrix. 

\section{Quantifying the impact of the choice of the redshift distribution}
\label{sec:pz_systematics}

\begin{figure}
	\centering
	\includegraphics[width= 8cm]{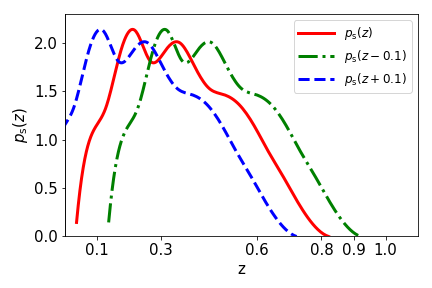}
	\caption{Comparison between our fiducial WISE redshift distribution $p_{\rm s}(z)$ and the ones obtained shifting it by $\Delta z= \pm0.1$.}
\label{fig:offset_dndz}
\end{figure}

\begin{figure*}
	\centering
	\includegraphics[width=18cm]{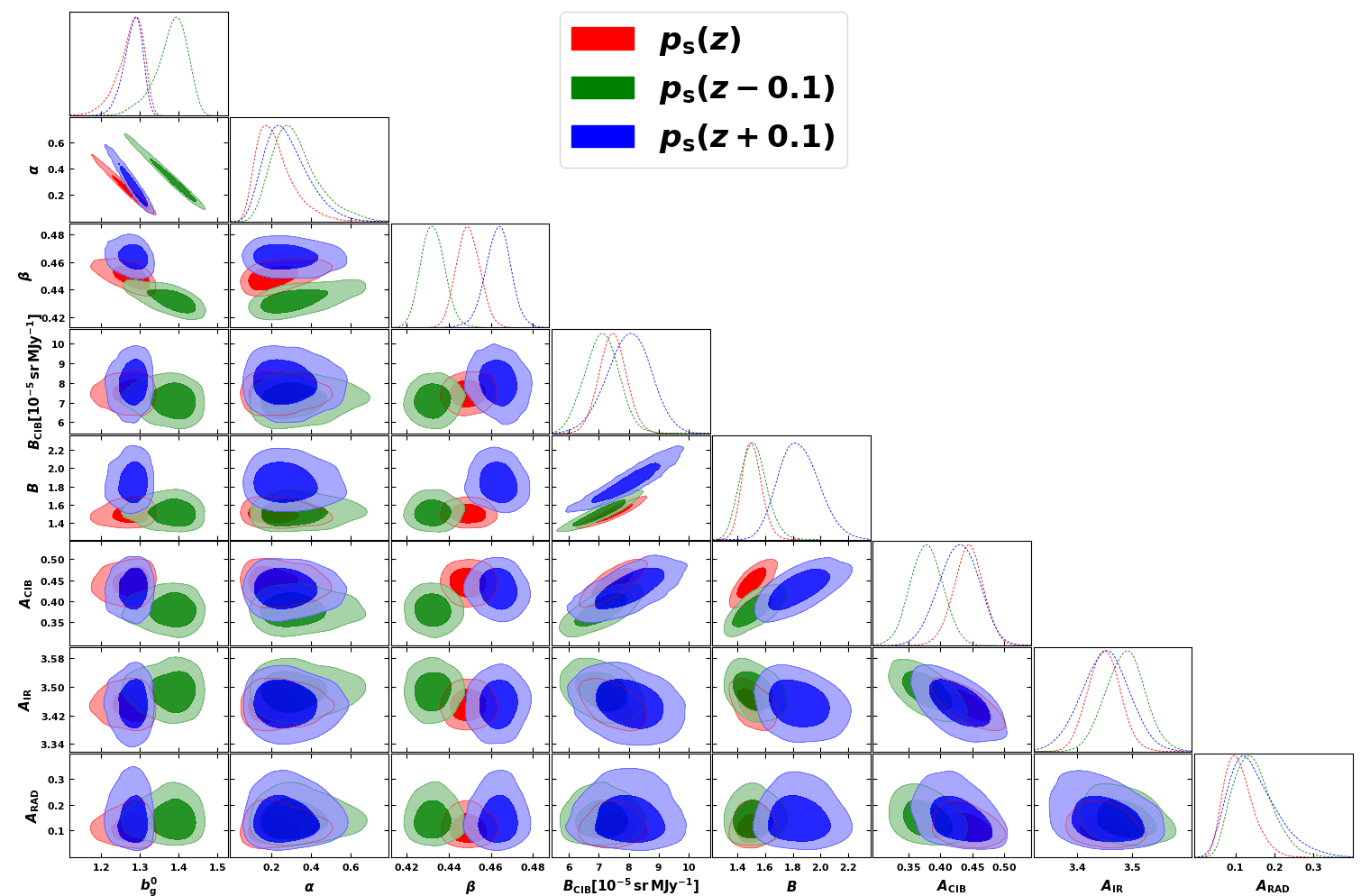}
	\caption{Posterior distributions on our model parameters for different estimation cases, corresponding to each of the WISE redshift 
	distributions plotted in Fig.~\ref{fig:offset_dndz}.}
	\label{fig:offset_posteriors}	
\end{figure*}

We show in this section some further details on the estimation of the systematic error bars based on different choices of the 
WISE redshift distribution. The offset $p_{\rm s}(z)$ distributions, compared with the original one, are shown in Fig.~\ref{fig:offset_dndz}, 
while the posterior distributions obtained repeating the parameter estimation procedure with each of them are shown in Fig.~\ref{fig:offset_posteriors}.
The best-fit parameters obtained in each case (which are used to estimate the systematic error component), are instead quoted in Table~\ref{tab:offset_estimates}.

\begin{table*}
\centering
\caption{Comparison of the constraints obtained on our model parameters using our fiducial WISE redshift distribution and the ones 
	shifted by $\Delta z= \pm0.1$. In each case we quote the best-fit estimate with the associated statistical error bars. The unit of $B_{\rm CIB}$ is $[10^{-5}\,{\rm sr}\,{\rm MJy}^{-1}]$.}
\label{tab:offset_estimates}		
\begin{tabular}{cccc} 		
\hline  
\hline 
Parameter     &  $p_{\rm s}(z)$ & $p_{\rm s}(z-0.1)$ &$p_{\rm s}(z+0.1)$ \\
\hline
$b_{\rm g}^0$ &  $1.28^{+0.03}_{-0.04}$& ${1.39}^{+0.04}_{-0.05}$& ${1.28}^ {+0.02}_{-0.03} $ \\

${\alpha}$ & $0.20^{+0.11}_{-0.07}$&${0.31}^{+0.14}_{-0.10}$&${0.26}^{+0.13}_{-0.09} $ \\

${\beta}$ &  $0.45{\pm 0.01} $ &${0.43}\pm{0.01}$ &${0.46}\pm {0.01}$\\

$B$ &  $1.50{\pm 0.07}$ &${1.51}^{+0.10}_{-0.09}$&${1.84}^{+0.15}_{-0.14} $\\

$B_{\rm CIB}$ & ${7.46}^{+0.45}_{-0.44}$ &${7.11}^{+0.56}_{-0.58}$&${ 8.05}^{+0.73}_{-0.76}$\\

$A_{\rm CIB}$ & $0.44{\pm 0.02}$ &${0.38}{\pm 0.03}$&${0.43}{\pm 0.03}$\\

$A_{\rm IR}$ & $3.45\pm{0.03}$&${3.49}^{+0.03}_{-0.04}$&${3.45}\pm{0.04}$ \\

$A_{\rm Rad}$ &  $0.10^{+0.04}_{-0.03}$ &${0.14}\pm{0.05}$&${0.14}^{+0.07}_{-0.05}$\\
\hline
\end{tabular}
\end{table*}

\bibliography{references}{}
\bibliographystyle{aasjournal}

\label{lastpage}
\end{document}